%% file: paper.tex
\documentclass[a4paper,showpacs]{iopart}
\usepackage{amsfonts}
\usepackage{graphicx}
\usepackage{color}
\usepackage{bm}

\include{macroz}

\newcommand{\TT}{{\cal T}}

\newcommand{\diag}{{\rm diag}}
\newcommand{\ndiag}{{\rm non-diag}}
\newcommand{\reg}{{\rm reg}}
\newcommand{\sing}{{\rm sing}}

\renewcommand{\x}{{\bf x}}
\renewcommand{\c}{{\rm c}}
\renewcommand{\o}{{\rm o}}

\newcommand{\Dhat}{\hat{D}}
\newcommand{\Ghat}{\hat{G}}
\newcommand{\GGh}{\hat{\cal G}}
\newcommand{\Kh}{\hat{K}}
\newcommand{\Ihat}{\hat{I}}

\newcommand{\Shat}{\hat{S}}
\newcommand{\That}{\hat{T}}
\newcommand{\Xhat}{\hat{X}}

\renewcommand{\phat}{\hat{p}}

\newcommand{\pnhat}{\hat{P}}

\renewcommand{\Th}{\hat{\cal S}}
\renewcommand{\T}{{\cal S}}

\begin{document}
\title{In-out decomposition of boundary integral equations}
\author{Stephen C. Creagh, Hanya Ben Hamdin and Gregor Tanner}
\address{School of Mathematical Sciences,
University of Nottingham, University Park, Nottingham, NG7~2RD, UK}
\begin{abstract} 
We propose a reformulation of the boundary integral equations for 
the Helmholtz equation in a domain in terms of incoming and outgoing
boundary waves.
We obtain transfer operator descriptions which are exact and thus incorporate 
features such as diffraction and evanescent coupling; 
these effects are absent in the well known semiclassical transfer operators in the 
sense of Bogomolny.
It has long been established that transfer operators are equivalent to the boundary integral 
approach within semiclassical approximation. 
Exact treatments have been restricted to specific boundary conditions 
(such as Dirichlet or Neumann). 
The approach we propose is independent of the boundary 
conditions, and in fact allows one to decouple entirely the problem
of propagating waves across the interior from the problem of reflecting 
waves at the boundary. 
As an application, we show how the decomposition
may be used to calculate Goos-H\"anchen shifts of ray dynamics in 
billiards with variable boundary conditions and for dielectric cavities.
\end{abstract}

\section{Introduction}
The transfer operator formalism proposed by Bogomolny 
 \cite{Bog} has offered a very powerful platform for the 
application of semiclassical methods to complex quantum and wave 
problems. 
It has long been recognised that, in 
the special case of cavity problems such as quantum 
billiards or dielectric resonators, the transfer operator is 
intimately connected with boundary integral methods, which provide
the most effective means of exact, numerical solution. Boasman 
established this connection explicitly in 
\cite{Boasman} for quantum billiards, recasting the boundary 
integral equations as the application of a transfer operator to the 
wavefunction on the boundary (in the case of Dirichlet boundary 
conditions) or to its normal derivative (in the case of Neumann 
boundary conditions). Subsequent development using Fredholm theory can
be found in \cite{GP1,GP2,HS,THS,ghosts}, particularly on the
Dirichlet case. If the boundary conditions are more general, 
however, it is less clear how the connection can be made exactly, 
although the approaches are easily related semiclassically (see
\cite{SieberRobin,Sieber3D,Blumel1,Blumel2}, for example). In this 
paper we propose that a decomposition of the wavefunction at the boundary
into incoming and outgoing components, defined in detail following a 
separation of the Green function into its regular and singular 
parts, provides a natural means of establishing this connection 
beyond leading semiclassical approximations.

Semiclassical methods originating in the field of quantum chaos have found widespread
application in classical wave problems  with complex 
or chaotic ray limits. In particular in the context of  vibro-acoustics \cite{Tan07} 
or for electromagnetic wave fields \cite{Hem12}, cavity problems with 
a variety of boundary conditions are prevalent. An efficient phase space 
flux method, the so-called Dynamical Energy Analysis (DEA),
has been developed recently, predicting wave energy transport 
through complex structures and domains based on trajectory 
calculations \cite{Tan09, Cha13}. A more explicit connection to boundary 
integral equations in terms of  transfer operator methods and incoming and outgoing 
waves is a key  ingredient for constructing hybrid DEA - wave methods 
\cite{Maks11}. This is especially valuable where semiclassically
higher-order mechanisms such as diffraction or evanescent coupling 
are important. In addition, one often encounters the situation where 
semiclassical ideas provide a simple qualitative description of the problem, but 
where the wavelength is too large to use them for reliable quantitative 
predictions. A semiclassically-motivated decomposition such as proposed in this 
paper can then lead to an efficient implementation of fully wave-based calculations. 
Our treatment of the Goos-H\"anchen shift 
is one example of this latter case; simplistic ray-dynamical models 
break down in the most important regions of phase space but nevertheless 
provide a very useful conceptual  basis on which to found more accurate 
calculations using, for example, the method proposed here.

\begin{figure}[h]\begin{center}
\includegraphics[scale=0.3]{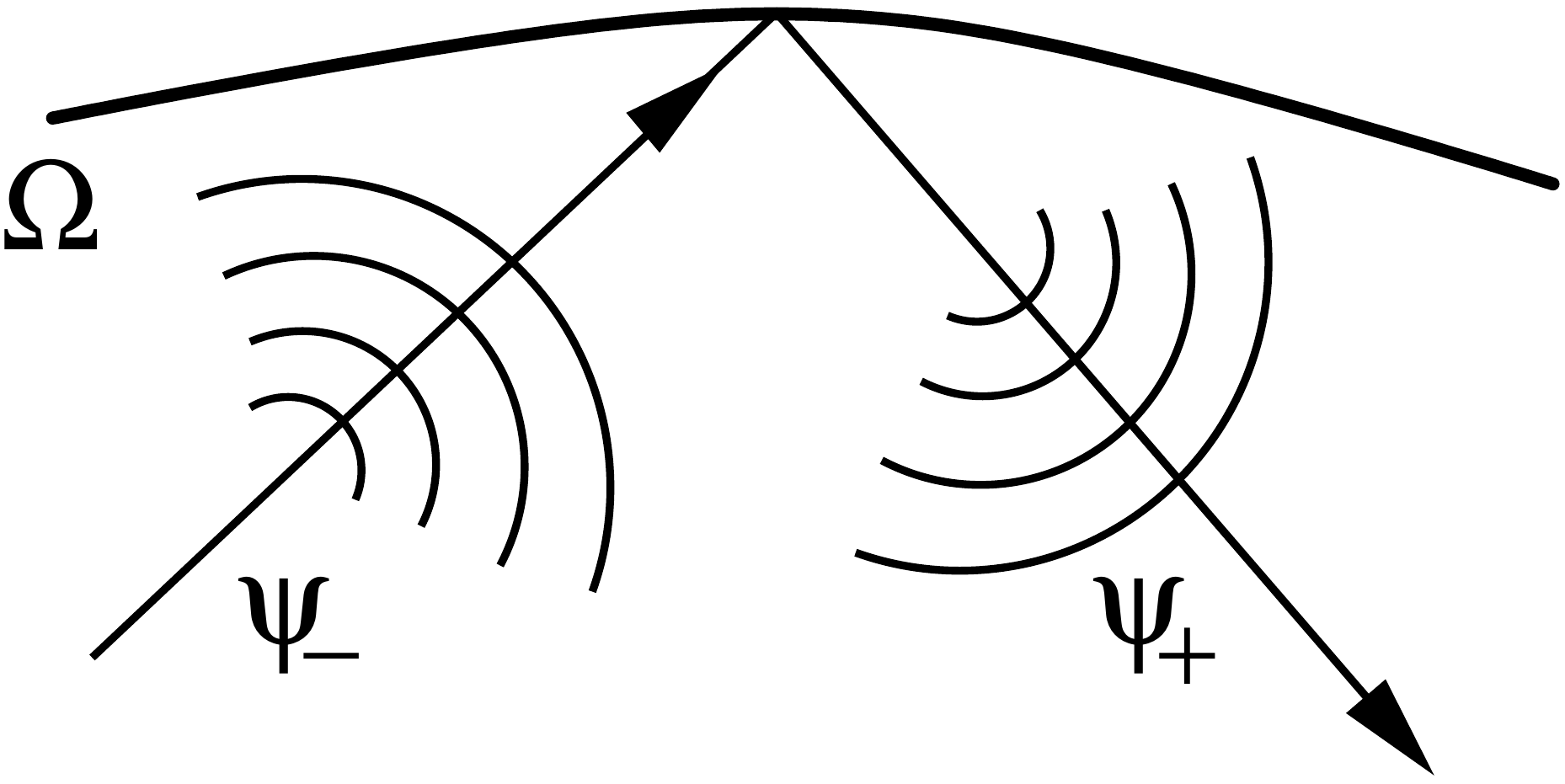}\end{center}
\caption{In this paper, $\psi_-(s)$ denotes the component of the boundary
solution corresponding to the part of the wave approaching the boundary 
from the interior of the domain $\Omega$ and $\psi_+(s)$ denotes the 
component leaving the boundary towards the interior.}\label{inoutfig}
\end{figure}

The main technical issue to be addressed in this paper is how to
decompose the wave solution at the boundary into a 
component approaching the boundary and a component leaving 
it (see Fig.~\ref{inoutfig}). This decomposition is automatically 
achieved in the context of semiclassical approximation when the 
wavefunction takes the form of an eikonal ansatz, where ray direction
allows us to select incoming and reflected wave components.
For problems which are not semiclassical, or where the solution
cannot easily be represented in eikonal form, the decomposition is less 
obvious. We propose 
that the outgoing wave $\psi_+(s)$ should be defined simply as the 
contribution to the boundary integral equation arising from an
appropriately defined singular 
part of the Green function. The motivation 
for this definition is that, just inside $\Omega$, the corresponding 
contribution to the boundary integral equation is dominated by the 
part of the boundary nearby and represents a wave emerging from the 
boundary towards the interior. This definition is consistent within
semiclassical approximation with the obvious decomposition according 
to the direction of rays. It also reproduces expected results in 
the one-dimensional case and in a sense our calculation amounts to a 
generalisation to higher dimensions of the scattering approach 
normally used on quantum graphs \cite{GSgraphreview,Kgraphreview}.

There remains some ambiguity, however, arising from the possibility
of defining the singular part of the Green function in different ways.
The simplest definition having the correct semiclassical limit is 
to define the singular part to correspond to the Green function obtained
by replacing the boundary locally by its tangent line or plane. 
The resulting decomposition can also be motivated by 
a transformation to momentum-represention of the solution along the 
boundary.  This provides a separation according to ray direction as is 
automatically achieved using an eikonal ansatz. We will refer to this as the 
{\em primitive decomposition}. 
The primitive decomposition is shown in \cite{Hanyathesis} to 
allow the treatment of diffractive effects from discontinuous boundary 
conditions, for example, or to provide a basis for the semiclassical 
coupling of transfer operators between multiple domains with different
local wave number.

The primitive decomposition displays singularities in momentum 
representation near the case of critical reflection where waves arrive 
at the boundary almost tangentially. Although one might still define
an associated transfer operator before semiclassical approximation,
such singularities will at least lead to issues such as slow 
convergence. The regime of near tangential angle of incidence 
is particularly important when describing leakage 
from dielectric resonators. Here the ``critical line'' in phase 
space, along which refractive escape switches on, plays an 
important role in emission patterns and decay rates. We therefore
propose a second separation of the Green function into singular 
and regular parts based on a splitting of the boundary 
integral into distinct contours in the complex plane, one of 
which captures the Green function's singularity. We call this 
second decomposition the {\em regularised decomposition}. It successfully 
accounts for the important effects of boundary curvature at critical 
reflection and remains well-behaved when waves arrive nearly 
tangentially. We examine the regularised decomposition in detail for the special 
case of circular cavities. 

We find in general 
that the boundary integral equations can be cast in the form
\begin{equation}\label{shiftop}
\psi_- = \Shat\psi_+,
\end{equation}
where the operator $\Shat$ depends on the geometry of the 
domain but is completely independent of the boundary conditions: 
in this paper we will refer to $\Shat$ as the \textit{shift operator}.
This equation is naturally interpreted as saying that $\psi_-$
is obtained by propagating the outgoing wave $\psi_+$ 
across the interior until it returns to the boundary as an incoming 
wave. For the case of circular cavities we provide
explicit, closed-form expressions for $\Shat$. 
We emphasise that this is achieved even in the 
presence of variable boundary conditions such as those treated
in \cite{BerryDennis,Dubphase}, where the problem as a 
whole is nonintegrable. We show that, closing the system by using 
the boundary conditions to relate the outgoing to the incoming 
component by a reflection operator $\rh$, 
we arrive at an 
overall transfer operator taking the form of a stroboscopic map 
$\That=\rh\Shat$. 
Semiclassical treatments provide a simple interpretation of $\rh$ 
as acting on rays by a Goos-H\"{a}nchen shift, so that the overall 
operator offers a simple explanation of the Goos-H\"{a}nchen-perturbed
ray dynamics of the type used in \cite{SH,elloops,nonHam,NonHam2} to treat 
dielectric cavities. Furthermore, the regularised decomposition allows
a more complete treatment of the case of critical reflection 
(or refraction), albeit in that case with a less simple interpretation 
in terms of rays.

At a formal level, the construction of a transfer operator as a map
between boundary functions is similar to the scattering-matrix 
approach developed by Smilansky and coworkers 
\cite{DoronSmilansky,RouvinezSmilansky,SchanzSmilansky} 
or the transfer-operator approach by Prosen \cite{Prosen1,Prosen3,Prosen4}.
Similar to our approach,  a wave solution is decomposed in those calculations into components 
passing in either direction through a surface of section cutting 
through the domain of interest or split into an incoming and outgoing
wave component at (parts of) the boundary of a cavity. 
The formalism by Smilansky et al and Prosen 
rely on being able to construct explicitly a scattering matrix for the regions 
lying to either side of a surface of section. They are thus most natural when 
decomposing the problem in hand into two 'half-problems' separated by 
the surface of section. Such a  treatment becomes less obvious when 
using as the section the full boundary of a closed cavity, the most 
natural surface of section when starting from boundary integral equations.
It is worth mentioning that the  transfer operator construction proposed by 
Prosen \cite{Prosen1,Prosen3,Prosen4} is similar to the 'primitive 
decomposition' introduced above. Prosen developed a more general 
approach dealing also with smooth potentials, but the restricted 
circumstances we assume here allows a simpler derivation and a 
generalisation to a regularised decomposition which deals 
better with the case of critical reflection, for example.

The emphasis in this paper is quite different, however. We argue
that, starting from the boundary integral equation, one can naturally
decompose the Green functions occurring there into singular and 
regular components that are closely related to the incoming
and outgoing component of the boundary wave function. The regularised
decomposition provides in particular a description which stays
regular at grazing incident angles and which makes it possible to distinguish 
propagating and evanescent (exponentially decaying) wave contributions
without explicitly referring to a (scattering) channel basis. 

The regularised decomposition approach applies in it simplest form to 
cavities that have convex, analytic boundaries. The boundary needs to 
be analytic so that the contour integration methods used to define 
singular and regular parts of the Green operators can be
defined. Assuming that the boundary is convex allows a more 
direct leading semiclassical description of the shift operator $\Shat$
in terms of the Green operators, without the need to account explicitly
for the cancellation of ghost orbits (see
\cite{ghosts}). Treatments of piecewise analytic boundaries, with a
more explicit account of corners, and of nonconvex billiards will be
the subject of future investigation.

We conclude this section by summarising the content of the paper. We begin in
Sec.~\ref{overviewsec} by establishing notation and outlining the 
main technical features of the calculation. The use of the
decomposition of Green operators into local and nonlocal parts to
define incoming and outgoing waves is illustrated in
Sec.~\ref{1dsec} for the one-dimensional case, where calculations are 
elementary. This is generalised to two-dimensional cavities in
Sec.~\ref{primsec} using the primitive decomposition. The regularised
decomposition, taking more explicit account of boundary curvature, is
treated in Sec.~\ref{regsec}. Finally some model two-dimensional cases are
examined in Secs.~\ref{egsec} and \ref{sec:dielcav}, providing in
particular explicit analytic descriptions of the shift operators for
the circle,  and conclusions are offered 
in Sec.~\ref{concsec}.

\section{Overview}\label{overviewsec}
We now summarise the main features of our calculation, concentrating 
on the case of two-dimensional cavities to fix notation. We emphasise 
that the main ideas in this section generalise to higher 
dimensions.

Let $\psi(\x)$ satisfy the Helmholtz equation
\[
-\nabla^2\psi = k^2\psi
\]
in the (two-dimensional) domain $\Omega$. We use the same symbol 
$\psi(s)$ to represent the restriction of the solution to the 
boundary $\partial\Omega$, on which $s$ is an arc-length coordinate, 
and denote the normal derivative by
\[
\mu(s) = \dydxv{\psi}{n}.
\]
Let $G_0(\x,\x';k)$ denote the free-space Green function satisfying
$(-\nabla^2-k^2)G_0(\x,\x';k)=\delta(\x-\x')$.
The boundary integral equation for $\psi$ may then be written formally as
\begin{eqnarray}\label{bie}
\psi&=&\Ghat_0\mu-\Ghat_1\psi,
\end{eqnarray}
where we denote the Green operators
\begin{eqnarray}
\label{defG0}\Ghat_0\mu (s)&=&
\lim_{\x\to s}\int_{\partial\Omega}G_0(\x,s';k)\mu(s')\d s'\\[3pt]
\label{defG1}\Ghat_1\psi(s)&=&
\lim_{\x\to s}\int_{\partial\Omega}\dydxv{G_0(\x,s';k)}{n'}\psi(s')\d s',
\end{eqnarray}
in which $\x$ denotes a generic point in the interior of $\Omega$ and
$s$ and $s'$ label points on the boundary.

\begin{figure}[!htp]
\centering
\includegraphics[scale=0.25]{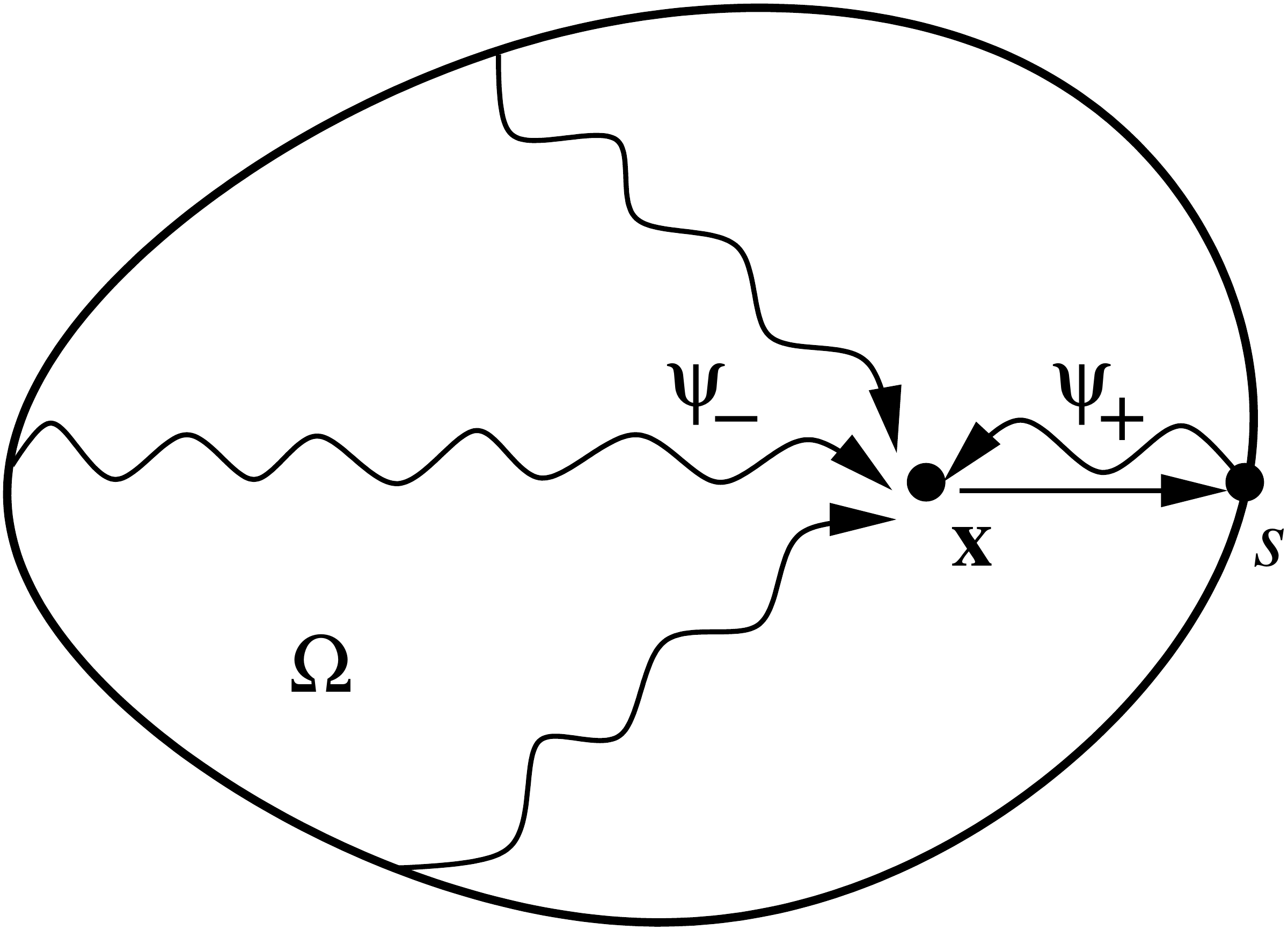}
\caption{The wave emerging from the part of the boundary near 
$\x$, as $\x\to\x(s)$, provides a natural definition of the outgoing 
component $\psi_+$. We define this more directly as the contribution
from the singular part of the Green function.
The regular remainder defining 
$\psi_-$ represents the contributions originating from the 
rest of the boundary, arriving at $s$ having passed through 
the interior.}
\label{2Dbeam}
\end{figure}

The key step in the transformation we propose 
is to perform a basis change
\[
\left(\begin{array}{c}\psi(s)\\\mu(s)\end{array}\right)\to
\left(\begin{array}{c}\psi_-(s)\\\psi_+(s)\end{array}\right)
\]
in which we replace the boundary wavefunction $\psi(s)$ and normal derivative
$\mu(s)$ 
by incoming and outgoing components $\psi_-(s)$
and $\psi_+(s)$, defined in the simplest implementation so that 
\begin{equation}\label{defpsipm2d}
\psi(s) = \psi_-(s) + \psi_+(s).
\end{equation}
Motivated by the schematic picture of Fig.~\ref{2Dbeam}, we argue 
that $\psi_+(s)$ is naturally defined as corresponding to the part of the
boundary integral in which $s'$ approaches $s$. This should capture in 
particular the singularity of the 2D Green function,
\begin{equation}\label{singularity}
G_0(s,s',k) \sim -\frac{1}{2\pi}\log(k|s-s'|)
\end{equation}
as $s'\to s$. We therefore separate 
the Green operators into ``singular'' and ``regular'' parts,  
\begin{equation}\label{singreg}
\Ghat_0=\Ghat^\sing_0+\Ghat^\reg_0,
\end{equation}
(with an analogous decomposition of $\Ghat_1$), where $\Ghat_0^\sing$
is an operator which should give the contribution to the 
boundary integral from $s'$ near $s$ and should in particular 
account for  the singularity 
(\ref{singularity}). We then define
\begin{equation}\label{singeq}
\psi_+ = \Ghat^\sing_0\mu-\Ghat^\sing_1\psi,
\end{equation}
and
\begin{equation}\label{regeq}
\psi_- = \Ghat^\reg_0\mu-\Ghat^\reg_1\psi.
\end{equation}
Note that, in the schematic picture of Fig.~\ref{2Dbeam}, $\psi_-(s)$
now collects the contributions to the boundary integral equation from 
the remainder of the boundary, arriving at $s$ having crossed the 
interior of the cavity. This is consistent with our requirement
that $\psi_-$ should represent the part of the solution that is incoming at 
the boundary.

We discuss in the coming sections the detailed calculation that arises
once the decomposition (\ref{singreg}) has been defined more explicitly.
Here we summarise the main conclusion, which is, that the boundary
integral equation (\ref{bie}), expressed in terms of $\psi_+$ and 
$\psi_-$, can be recast formally as equation (\ref{shiftop}).
We emphasise that this equation derives entirely from the boundary integral equation
and has nothing to do with the boundary conditions. For a given domain 
$\Omega$, we get the same shift operator $\Shat$ in  (\ref{shiftop}) 
regardless of whether Dirichlet, Neumann or any other boundary condition is
imposed.

The boundary conditions provide us with a second, independent 
relationship between $\psi_-$ and $\psi_+$. Suppose that the 
boundary conditions are prescribed in the form of a linear relationship 
between $\psi$ and $\mu$ on $\partial\Omega$. (This assumption must be 
modified for the treatment of dielectric and other problems 
where coupling to an exterior solution is involved, as described in 
Sec.\ \ref{sec:dielcav}). 
We re-express this in terms of the functions $\psi_-(s)$ and
$\psi_+(s)$ as
\[
\psi_+ = \rh\psi_-
\]
where we call $\rh$ the \textit{reflection operator}, for obvious reasons.
Combined with (\ref{shiftop}), we find
\[
\psi_+ = \That\psi_+,
\]
where 
\begin{equation}\label{defThat}
\That = \rh\Shat
\end{equation}
is the \textit{transfer operator} for the problem as a whole. 
Eigenvalues are found by solving the secular equation
\[
\det(\Ih-\That(k)) = 0.
\]
We find indeed that, when semiclassical approximations are used for
$\Shat$ and $\rh$, we recover the usual transfer operator expected 
from \cite{Bog} (up to a simple conjugation of the operators 
involved: see \ref{Appsc}).

Note that the transfer operator $\That$ is presented as a stroboscopic 
map, composing the shift operator $\Shat$, determined entirely by the
shape of $\Omega$, with the reflection operator $\rh$, accounting for
boundary conditions. In semiclassical approximation, $\Shat$ is determined 
by the usual ray dynamics assuming specular reflection at each bounce.
The reflection operator can be accounted for in this picture simply by 
adding appropriate phases as rays are reflected from the boundary.
An alternative point of view, which has been found to be fruitful in the 
treatment of dielectric problems, for example, is to regard the
variable reflection phases inherent in $\rh$ as adding a perturbation 
to the usual ray dynamics in the form of a Goos-H\"anchen shift 
\cite{SH,elloops,nonHam,NonHam2,GFIEEE,GFQE}. In our 
approach, this is easily understood from the stroboscopic nature 
of the full operator $\That$. Writing $\rh$ formally as an exponential
\begin{equation}\label{reflgen}
\rh = \e^{-\rmi k\hh} ,
\end{equation}
and using an analogy with quantum mechanics in which $k$ plays the 
role of $1/\hbar$, we can interpret $\hh$ as the generator of $\rh$.
The corresponding effect on rays is obtained simply by replacing $\hh$
by a symbol $h(s,p)$ on the boundary phase space and using this as a 
classical Hamiltonian. Note that, because the reflection phases are
formally of $O(1)$ in an eikonal expansion in $1/k$, the Hamiltonian
$h(s,p)$ so defined is small and therefore only perturbs the main 
ray dynamics (and that is why we are equally entitled to leave the ray 
dynamics alone and incorporate reflection phases instead in the 
next-to-leading, amplitude-transport phase of the eikonal expansion 
\cite{usell}).

\section{The one-dimensional case and quantum graphs}\label{1dsec}
In order to motivate equations (\ref{singeq}) and (\ref{regeq}) further
as a means of decomposing the boundary solution into incoming and outgoing components,
we consider now the one-dimensional case. The most direct analogue
leads very simply to the standard picture one uses for the wave dynamics  
on quantum graphs \cite{GSgraphreview}.

\begin{figure}[!htp]
\centering
\includegraphics[scale=0.25]{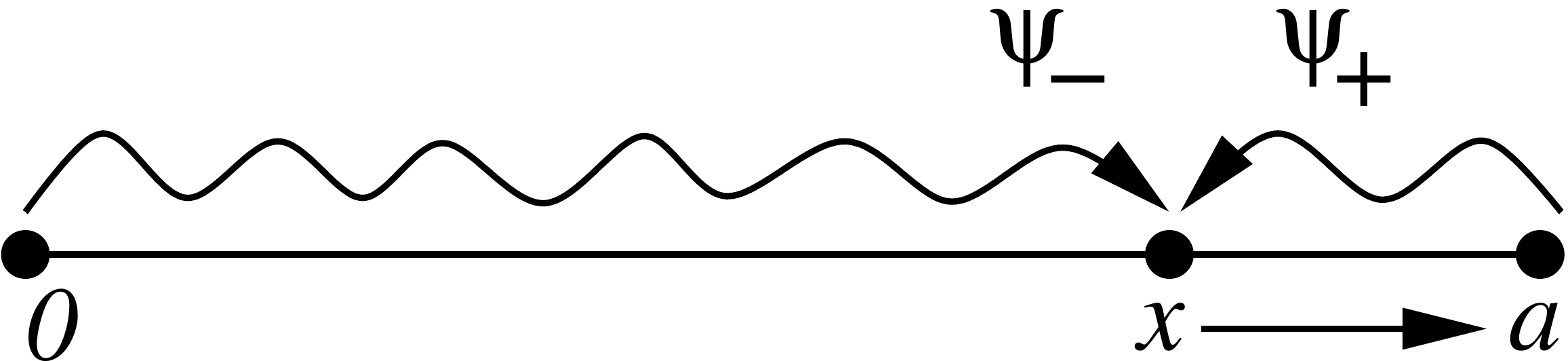}
\caption{A schematic illustration is given of the diagonal and nondiagonal
contributions to the boundary integral equation in the one-dimensional case. 
As the interior point 
$x$ approaches the boundary point $a$, the off-diagonal contribution $\psi_-$
to (\ref{bie}) represents a wave originating at $0$ and moving towards the 
selected boundary point $a$. This diagonal contribution $\psi_+$ 
represents a wave emerging from $a$ and moving away from the boundary.}
\label{1Dbeam}
\end{figure}

Consider first the case of an interval $(0,a)$, illustrated in 
Fig.~\ref{1Dbeam}. Here the boundary solution can be represented by 
the two-component vectors
\[
\psi = \spinor{\psi(0)}{\psi(a)}
\qquad\mbox{and}\qquad
\mu = \spinor{\mu(0)}{\mu(a)},
\]
while, from the one-dimensional Green function
\begin{equation}\label{G01D}
G_0(x,x';k) = \frac{\rmi}{2k}\e^{\rmi k|x-x'|},
\end{equation}
one finds that the Green operators can be represented by the $2\times 2$
matrices
\begin{equation}\label{G01d}
G_0 = 
\frac{\rmi}{2k}\left(\begin{array}{cc}
       1 & \rme^{\rmi ka} \\[0.3em]
       \rme^{\rmi ka} & 1 
     \end{array}\right)
=\frac{\rmi}{2k}\left( I+S\right)
\end{equation}
and
\begin{equation}\label{G11d}
G_1 = -\frac{1}{2}\left(\begin{array}{cc}
       1 & \rme^{\rmi ka} \\[0.3em]
       \rme^{\rmi ka} & 1 
     \end{array}\right)
=-\frac{1}{2}\left( I+S\right),
\end{equation}
where $I$ is the identity matrix and
\begin{equation}\label{defT1d}
S = \left(\begin{array}{cc}
       0 & \rme^{\rmi ka} \\[0.3em]
       \rme^{\rmi ka} & 0
     \end{array}\right)
\end{equation}
will be found to play the role of the shift operator.

In this one-dimensional case, instead of separating the Green function 
into a singular and a regular part, as in (\ref{singreg}), we 
separate the diagonal from the nondiagonal part
\begin{equation}
G_0=G^\diag_0+G^\ndiag_0.
\end{equation}
That is, the role played in higher dimensions by the singular part 
(as $s'\to s$)
of the Green function is in one dimension taken over by the diagonal part (in which 
$s'=s$). 
Then, defining
\begin{equation}\label{psiplus1d}
\psi_+ = G^\diag_0\mu-G^\diag_1\psi
=\frac{\rmi}{2k}\mu+\frac{1}{2}\psi,
\end{equation}
and
\begin{equation}\label{psiminus1d}
\psi_- = G^\ndiag_0\mu-G^\ndiag_1\psi
=\frac{\rmi}{2k}S\mu+\frac{1}{2}S\psi,
\end{equation}
in analogy with (\ref{singeq}) and (\ref{regeq}), we find
that the boundary integral equation (\ref{bie}) becomes
\[
\psi_- = S\psi_+.
\]
Note that in the case of quantum graphs we simply apply the same 
calculation independently to each bond and find an overall shift 
operator with diagonal blocks of the form (\ref{defT1d}). The problem
is closed by applying boundary conditions at vertices to provide
scattering operators that generalise $\rh$ in (\ref{defThat}) and
couple the bond-labelled blocks of $\Shat$.

\section{The primitive decomposition}\label{primsec}
\begin{figure}[!htp]
\centering
\includegraphics[scale=0.25]{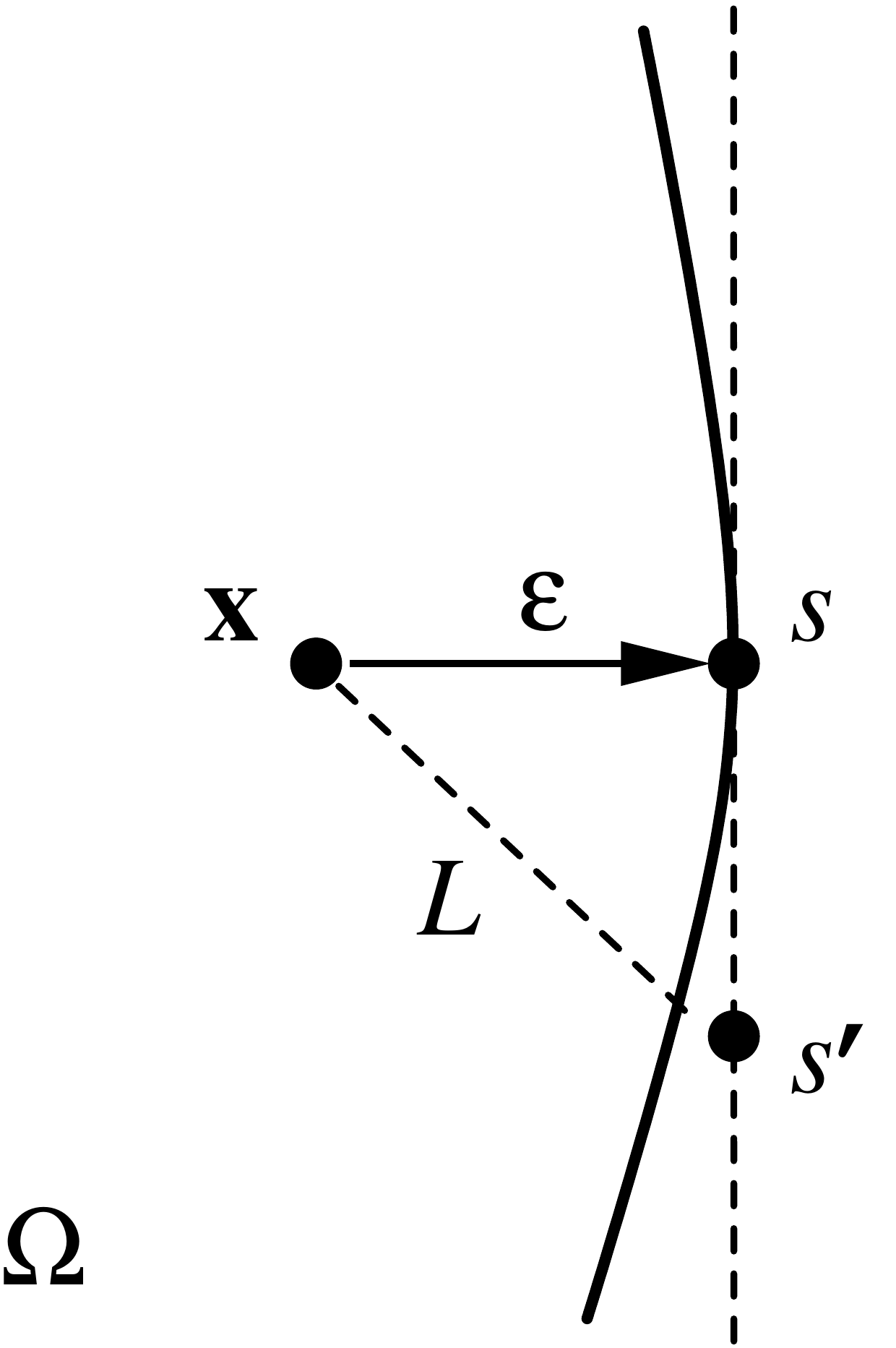}
\caption{Illustration of the local replacement of the boundary (curved 
solid line) with  the tangent (vertical dashed line) at $s$.}
\label{Sing}
\end{figure}

The simplest decomposition in the two-dimensional case is
obtained by defining the singular part of the Green operator
to be the result of replacing the boundary locally by its tangent 
line, as illustrated in Fig.~\ref{Sing}.

Before taking the limit, the kernel of the full Green operator entering
(\ref{defG0}) takes in 2D the form
\[
G_0(\x, s'; k) 
=\frac{\rmi}{4} H_0(kL(\x,s')),
\]
where $L(\x,s')$ denotes the distance between the generic interior point
$\x$ and the boundary point $s'$ and $H_0(z)$ is the outgoing Hankel 
function. If the boundary is replaced by the 
tangent line at $s$, (see Fig.~\ref{Sing}), this distance can be written
\[
L(\x,s')\sim\sqrt{(s-s')^2+\eps^2}, 
\]
where $\eps$ denotes the normal distance from $\x$ to $s$. 
We define a corresponding singular component of the 
Green operator by
\begin{equation}\label{defGsingint}
\Ghat_0^\sing\mu(s) = \lim_{\eps\to 0}
\frac{\rmi}{4}\int_{-\infty}^\infty H_0\left(k\sqrt{(s-s')^2+
\eps^2}\right)\mu(s')\rmd s'.
\end{equation}
The boundary function $\mu(s)$ entering the integrand in (\ref{defGsingint})
and in analogous equations below is here understood as being the periodic 
extension of the true boundary function onto the real line 
with period equal to the perimeter of $\partial\Omega$. 
(This is understood to hold both for open
and closed boundaries in order to unify notation.)

This operator takes a remarkably simple form when expressed in a Fourier basis
(or, in quantum mechanical language, a momentum representation):
\begin{equation}\label{defGsing}
\Ghat^\sing_0=\frac{\rmi}{2k\pnhat} ,
\end{equation}
where 
\begin{equation}
\pnhat=\sqrt{1-\phat^2}
\label{pnOP} 
\end{equation}
formally defines a normal momentum operator and
\begin{equation*}
\phat=\frac{1}{\rmi k}\frac{\partial}{\partial s}
\end{equation*}
denotes a tangential momentum operator.
More precisely, the action of the operator 
$\pnhat$ on a function $\psi(s)$ is defined with respect to the
Fourier transform 
\begin{equation}\label{defft}
\varphi(p) = 
\int_{-\infty}^\infty\rme^{-\rmi kps}
\psi(s)\rmd s
\end{equation}
to be 
\begin{eqnarray}
\pnhat\psi(s)
&=& \frac{k}{2\pi}
\int_{-\infty}^{\infty}\rme^{\rmi kps}\sqrt{1-p^2}\,\varphi(p)
\rmd p,
\label{oppn}
\end{eqnarray}
where integration is over a contour in the complex $p$-plane that 
passes above the branch singularity $p=-1$ and below the branch 
singularity at $p=1$. 
The singular part of the operator $\Ghat_1$ is even simpler:
\begin{equation}\label{defG1sing}
\Ghat_1^\sing = -\frac{1}{2}\Ihat,
\end{equation}
where $\Ihat$ is the identity operator. These relations are derived explicitly 
in \ref{SingpartG} and reconfirm the well-known jump 
condition important to boundary integral methods more generally.

The outgoing wave defined in (\ref{singeq}) can now be written
\[
\psi_+ = \frac{\rmi}{2k} \frac{1}{\pnhat} \mu + \frac{1}{2}\psi,
\]
providing a two-dimensional generalisation of (\ref{psiplus1d}).
Together with the condition $\psi=\psi_++\psi_-$, this implies
\[
\mu = \rmi k\pnhat\left(\psi_--\psi_+\right),
\]
which is what one might expect intuitively on the basis of
a Fourier representation of the solution near the boundary: 
a boundary solution $\psi(s)=\e^{\rmi kps}$ extends to an interior solution
$\psi_\pm(\x)\simeq\exp[{\rmi k(ps\mp\sqrt{1-p^2}n)}]$ near the boundary, where
$n$ represents a normal coordinate increasing towards the interior
 and the sign is determined by whether the
extended wave moves towards or away from the boundary. Taking a derivative 
with respect to $n$ then gives the form for $\mu$ written above.

To complete the calculation, and motivated by the one-dimensional
equations (\ref{G01d}) and (\ref{G11d}), 
let $\Shat_0$ and $\Shat_1$ be {\em defined} 
by
\begin{equation}\label{defT0}
\Gh_0 = \frac{\rmi}{2k}\left(\Ihat + \Shat_0\right)\frac{1}{\pnhat}
\end{equation}
and
\begin{equation}\label{defT1}
\Gh_1 = -\frac{1}{2}\left(\Ihat + \Shat_1\right)
\end{equation}
respectively. Then
\begin{eqnarray*}
\psi_- &\equiv& \Gh_0^\reg\mu - \Gh_1^\reg\psi
\\
&=&\frac{1}{2} \Shat_0\left(\psi_+-\psi_-\right) 
+ \frac{1}{2}\Shat_1\left(\psi_++\psi_-\right)
\end{eqnarray*}
and so
\begin{equation}\label{preT}
\left(\Ihat+\frac{1}{2}(\Shat_0-\Shat_1)\right)\psi_-
 = \frac{1}{2}\left(\Shat_0+\Shat_1\right)\psi_+,
\end{equation}
from which $\psi_+=\Shat\psi_-$, where,
\begin{equation}\label{defT}
\Shat = \left(\Ihat+\frac{1}{2}(\Shat_0-\Shat_1)\right)^{-1}
\frac{1}{2}\left(\Shat_0+\Shat_1\right)
\end{equation}
defines $\Shat$ formally.
Note that, although their one-dimensional analogues in
(\ref{G01d}) and (\ref{G11d}) are equal, the operators $\Shat_0$ and 
$\Shat_1$ are distinct. In the case of convex domains, they do,
however, have the same leading approximation 
within semiclassical expansion \cite{Hanyathesis}, so that
\begin{equation}\label{Ssequal}
\Shat\approx\Shat_0\approx\Shat_1.
\end{equation}
The relevant semiclassical expressions are described more explicitly
in \ref{Appsc}. 
Note also that, although (\ref{defT}) requires an operator inversion,
the operator to be inverted is close to the identity in semiclassical
approximation for convex cavities. One can then easily incorporate the inversion
into higher-order approximations in that case.

\section{Regularised decomposition for analytic boundaries}\label{regsec}
The primitive decomposition leads to singularities in momentum
representation around $p=\pm 1$ that are unphysical for curved boundaries
(see the discussion in Sec.~\ref{robsec} around Fig.~\ref{psiprimfigs}, for example).
This singular behaviour can be removed by taking account of boundary 
curvature more explicitly. 
We describe in this section an alternative
decomposition which achieves this goal (in two dimensions). 
As a concrete example, we will treat circular cavities in detail in Sec.\
\ref{egsec}, for which all the relevant operators can
be fully understood analytically. We emphasise, however, that 
much of the discussion generalises to other geometries
with analytic boundaries.

\subsection{Decomposition by integration contour}\label{contoursec}

To motivate the construction, let us examine the action of the 
Green operator $\Gh_0$, on the boundary plane wave
\[
\mu_p(s) = \e^{\rmi kps}.
\]
In asymptotic analysis it is natural to treat the resulting integral
\begin{equation}\label{defG0int}
\Gh_0\mu_p(s) = \frac{\rmi}{4}\int_{s}^{s+\ell}H_0(kL(s,s'))\e^{\rmi kps'}\d s',
\end{equation}
in which $\ell$ denotes the length of the perimeter, using the method 
of stationary phase and, where appropriate, exploiting 
the asymptotic representation of the Hankel function, which begins so that
\[
G_0(s,s';k)\simeq \frac{\rmi}{2}
\frac{\e^{\rmi kL(s,s')}}{\sqrt{2\pi\rmi kL(s,s')}}.
\] 
One can identify in particular two kinds of 
contribution: those from the endpoints at $s'=s$ and $s'=s+\ell$,
and those from saddle points.
At leading order, the asymptotic contribution of endpoints is 
obtained by Taylor-expanding the chord-length function and truncating
at the linear term, so that we use 
\[
L(s,s') = \left \{
\begin{array}{ll}
|s-s'| + O\left((s-s')^3\right),\\
|s-s'+\ell| + O\left((\ell - s+s')^3\right),
\end{array}
\right.
\]
respectively for the two endpoints. We now point out that this
truncated length 
function is precisely the 
length function (following the limit $\eps\to 0$) of the tangent line used in the previous section
to define $\Gh_0^\sing$. That is, the primitive definition of 
$\Gh_0^\sing\mu_p(s)$ amounts essentially to a separating out of the 
leading-order endpoint contributions from the integral in 
(\ref{defG0int}).

\begin{figure}[!htp]
\centering
\includegraphics[scale=0.19]{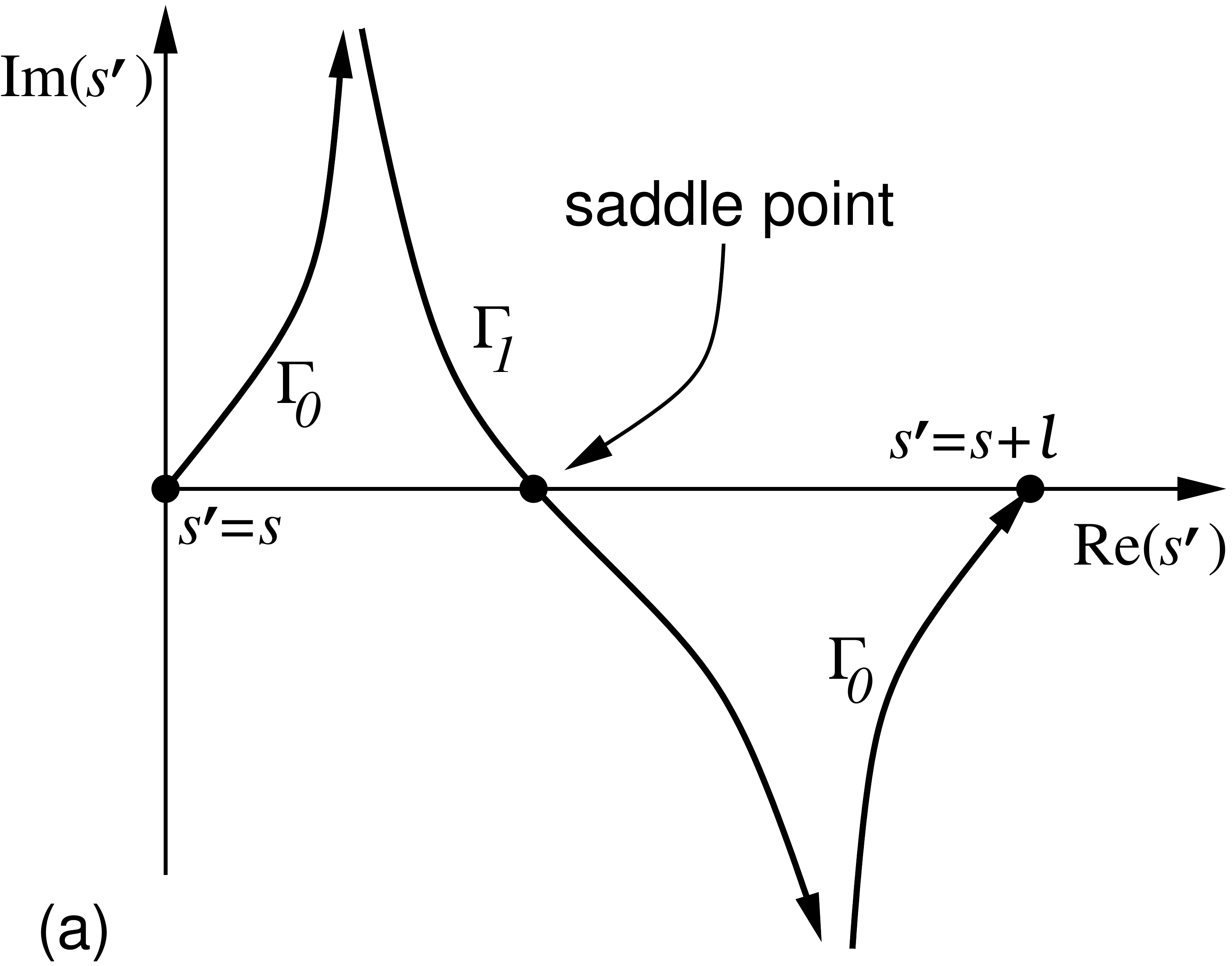}\hspace{12pt}
\includegraphics[scale=0.19]{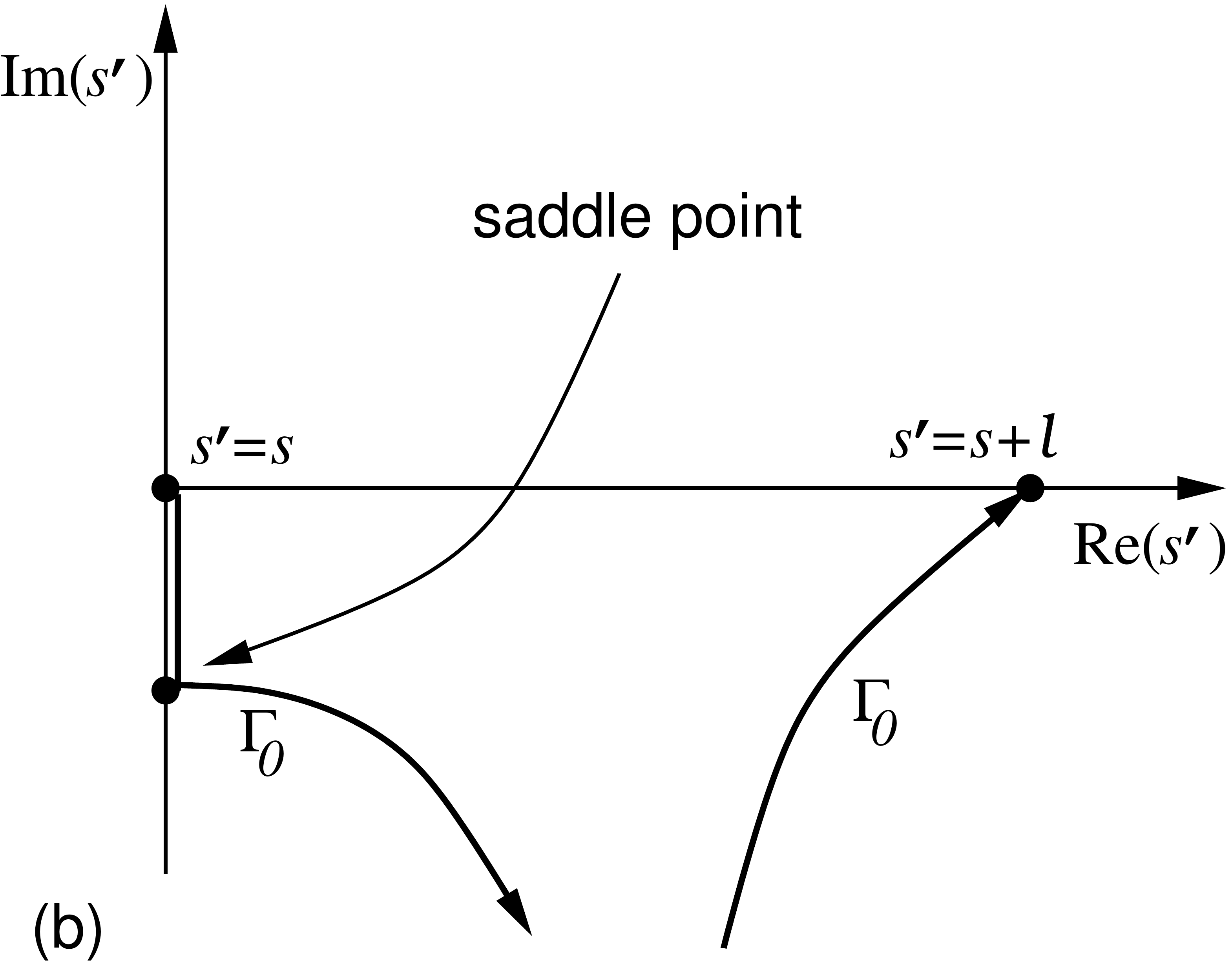}
\caption{The contours used to define the regularised decomposition
are illustrated schematically: in (a) for the case $p^2<1$ and in (b) for the case $p^2>1$.}
\label{contoursfig}
\end{figure}

This suggests that a better definition of $\Gh_0^\sing$
might be obtained by including higher-order corrections to the 
leading endpoint contribution already used. This would account 
in particular for curvature, which enters the Taylor 
series for $L(s,s')$ at third order. We assume in this 
section that the boundary is analytic. Then 
a systematic treatment of the full asymptotic expansion 
originating with the endpoints is obtained by promoting the 
integral in (\ref{defG0int}) to the status of a contour integral 
in the complex $s'$-plane and then isolating the contour component $\Gamma_0$
emerging from the endpoints and following a path of steepest descents. 
For appropriate test functions we might then define, simply,
\begin{equation}\label{singcont}
\Gh_0^\sing\mu(s) = \lim_{\x\to s}
\frac{\rmi}{4}\int_{\Gamma_0}H_0(kL(\x,s'))\mu(s')\d s'.
\end{equation}
This definition has the advantage of automatically offering the 
systematic asymptotic expansion of the endpoint contribution 
while also permitting, in principle, exact analysis.
In practice, this definition is complicated by the fact 
that the topology of $\Gamma_0$ depends in general on the test 
function. In particular, for the case of plane waves 
$\mu_p(s)$,  $\Gamma_0$ may take one of two topologically distinct routes, depending 
on whether $p^2<1$ or $p^2>1$ as illustrated in 
Fig.~\ref{contoursfig} for the case of the circle.

The contour for propagating waves with $p^2<1$
separates into a two-part endpoint component $\Gamma_0$ --- one part 
starting at $s'=0$ and ascending into the upper-half plane 
and the other part ascending from the lower-half plane and ending at 
$s'=s+\ell$ --- and a separate 
component $\Gamma_1$ which completes the contour and caters for real 
saddle points corresponding to the rays passing from $s'$ to $s$ across 
the interior. Note that such propagating waves correspond in the 
language of the scattering approach 
\cite{DoronSmilansky,RouvinezSmilansky,SchanzSmilansky,Prosen1,Prosen3,Prosen4}
 to open modes,
and to reflect this we let $\H_\o$ denote the subspace of boundary 
functions defined by $p^2<1$ in momentum representation. 

Thus, we {\em define} $\Gh_0^\sing$ on $\H_\o$ by (\ref{singcont})
and by fixing the contour $\Gamma_0$ to 
be the lines ascending vertically from $s'=s$ to the 
upper-half plane and ascending vertically from the lower-half plane 
to $s'=s+\ell$. Note that by fixing $\Gamma_0$ to rise vertically
it will in general deviate from the strict steepest-descent 
contour but will give the same result up to the exponentially small 
differences. These may arise when the respective contours go 
around singularities on different sides. Using a fixed contour has 
the advantage of letting us define the action $\Gh_0^\sing$ on $\H_\o$ 
without reference to a specific basis within that subspace.

If $\Gh_0^\sing$ is defined by (\ref{singcont}), the action of 
$\Gh_0^\reg$ on $\H_\o$ is then defined by
\begin{equation}\label{regcont}
\Gh_0^\reg\mu(s) = \lim_{\x\to s}
\frac{\rmi}{4}\int_{\Gamma_1}H_0(kL(\x,s'))\mu(s')\d s',
\end{equation}
where $\Gamma_1$ is now also fixed and understood to 
descend vertically from the upper-half plane 
to $s'=s$, to move across the real axis to $s'=s+\ell$ and then descend 
vertically into the lower half plane. Note that $\Gamma_1$ can 
be deformed so that it avoids entirely the endpoints of the
complete integral and is dominated in asymptotic analysis by
saddle points corresponding to interior-crossing rays (and potentially
their complex generalisations). We reiterate 
that at leading order in asymptotic analysis these definitions 
reproduce the action of the primitive decomposition defined in the
previous section.

For the subspace $\H_\c$ spanned by closed modes with $p^2>1$, 
asymptotic analysis uses a different contour decomposition, reflecting the
movement of contributing saddle points into the complex plane, directly below 
$s'=s$ (for $p<-1$) or above $s'=s+\ell$ (for $p>1$). 
In this case, the decomposition of 
contours used to define $\Gh_0^\sing$ and $\Gh_0^\reg$ is not
as obviously defined, suggesting that there are alternative, 
a priori equally sensible decompositions of closed boundary modes 
into incoming 
and outgoing components. In fact, the asymptotic form of the integral 
(\ref{singcont}) is in the middle of a Stokes transition when $p^2>1$ is real.
Illustrated in Fig.~\ref{contoursfig}(b)
is the contour decomposition for a test function with $p<-1$ in 
the circle. Here the endpoint contour segment $\Gamma_0$ descends vertically 
from $s'=s$ and runs into a saddle. If $p$ moves to either side 
of the real axis, $\Gamma_0$ passes to one side or other of the saddle. We may then
either include the saddle separately, by including a contour 
component $\Gamma_1$ running over it completely (see Fig.\ 
\ref{contoursbisfig} in  \ref{Appalt}) or miss it entirely. 
We choose the latter option, which amounts to
choosing $\Gamma_1=0$ and letting $\Gamma_0$ account for the entire contour,
as in Fig.~\ref{contoursfig}(b). This implies the choice 
$\Gh_0^\reg\H_\c=0$. See \ref{Appalt} for discussion of the
alternative convention.

The Green operator $\Gh_1$ could be decomposed analogously
by separating the appropriate boundary integral
using the same contour components. 
We note that, when taking the limit $\x\to s$ in this approach, a 
jump condition analogous to (\ref{defG1sing}) is satisfied, but that
there remains a nontrivial correction:
\begin{equation} \label{G1full}
\Gh_1^\sing = -\frac{1}{2}\Ihat + \Kh_1^{\sing},
\end{equation}
where 
\begin{equation}\label{G1int}
\Kh_1^{\sing}\mu(s) = 
\frac{\rmi k}{4}\int_{\Gamma_0} 
\cos \alpha'\,H_0'(kL(s,s'))\mu(s')\d s'
\end{equation}
is obtained by putting $\x$ directly onto 
the boundary and $\alpha'$ is an incidence angle defined 
in Fig.~\ref{chord}.

So far our discussion has suggested that each of
$\Gh_0$ and $\Gh_1$ might be decomposed individually by 
separation of contours. This approach has the advantage of being conceptually 
simple while allowing systematic asymptotic approximation 
beyond the leading primitive terms, and also allowing exact
analysis in principle. As will be shown in Sec.~\ref{unishift} below, 
it may be advantageous in practice to adopt a variation of this 
approach. In particular, the calculation of the shift operator $\Shat$ 
is greatly simplified if the identity
\begin{equation}\label{G1bis}
\Gh_1^\reg\left(\Gh_1^\sing\right)^{-1}=
\Gh_0^\reg\left(\Gh_0^\sing\right)^{-1}
\end{equation}
is satisfied. Note that this identity holds for convex domains at leading order 
for any of the decompositions we consider in this 
paper (including the primitive decomposition of 
section~\ref{primsec}). It is also shown in section~\ref{circsec}
that the definitions of $\Gh_0^\sing$ and $\Gh_1^\sing$ so far 
given yield the identity (\ref{G1bis}) exactly in the case of 
the circle. 

More generally, an alternative approach is to guarantee that
(\ref{G1bis}) holds
by choosing one of the following:
\begin{itemize}
\item[\textbf{A}] let (\ref{singcont}) define $\Gh_0^\sing$ and let 
$\Gh_1^\sing$ then be defined by the constraint (\ref{G1bis}) or
\item[\textbf{B}] let (\ref{G1full}) define $\Gh_1^\sing$ and let 
$\Gh_0^\sing$ then be defined by the constraint (\ref{G1bis}).
\end{itemize}
Because (\ref{G1bis}) holds in any case at leading order (for convex domains), higher 
corrections can be systematically included in an asymptotic 
analysis of this alternative approach. In this context, option 
\textbf{B} in particular is attractive because $\Gh_1^\sing$
has the simple leading approximation $\Gh_1^\sing\simeq-\Ihat/2$.

Finally, we note that if the length function $L(s,s')$ suffers 
singularities in 
the complex plane (in addition to those at $s'=s$ and 
$s'=s+\ell$), the function $\Gh_0^\sing\mu(s)$ defined 
by (\ref{singcont}) may exhibit discontinuities whenever these 
singularities cross $\Gamma_0$ (as $s$ is varied). 
Any such discontinuities will be small, however, as the integrand 
decays exponentially along $\Gamma_0$. In practical terms, such 
singularities can be circumvented by not allowing the contour segments
to extend indefinitely into the complex plane; that is, we may 
truncate $\Gamma_0$ (and therefore $\Gamma_1$) at a finite 
distance from the real axis, so that passing singularities are 
missed. Note that the detailed decomposition will then depend on 
the heights at which the contours are truncated, but only to the extent of
exponentially small corrections in asymptotic analysis. For the case 
of circular cavities described more fully in section \ref{egsec}, there are 
no such difficulties and we can give a complete analytical description of 
the resulting decomposition.

\subsection{Shift operator for the regularised decomposition}\label{unishift}
Let us now outline how a decomposition of the Green operators $\Gh_0$ and
$\Gh_1$ into singular and regular parts leads to a restatement of the 
boundary integral equations in terms of a shift operator. We do this 
initially without making any particular assumptions about the 
detailed properties of $\Gh_{0,1}^\sing$ and $\Gh_{0,1}^\reg$ and then 
point to the simplified results that may be obtained if explicit 
properties such as (\ref{G1bis}) are assumed. 

We start, in general terms, with equations (\ref{singeq}) and 
(\ref{regeq}) defining the outgoing and incoming components of 
the boundary solution. From (\ref{singeq}), we may eliminate 
\[
\mu = \left(\Gh_0^\sing\right)^{-1}
\left(\psi_++\Gh_1^\reg\psi\right)
\]
and, using also $\psi=\psi_++\psi_-$, we then find from (\ref{regeq}) that
\[
\left(\Ihat+\Gh_1^\reg-\Gh_0^\reg
\left(\Gh_0^\sing\right)^{-1}\Gh_1^\sing\right)\psi_-\qquad
\]\begin{equation}\label{genscat}\qquad
= \left(\Gh_0^\reg\left(\Gh_0^\sing\right)^{-1}-\Gh_1^\reg+\Gh_0^\reg
\left(\Gh_0^\sing\right)^{-1}
\Gh_1^\sing\right)\psi_+.
\end{equation}
Note that the primitive version (\ref{preT}) is obtained as a 
special case of this equation.

If we now also assume that the singular components
$\Gh_0^\sing$ and $\Gh_1^\sing$ are chosen, so that
(\ref{G1bis}) holds, this equation collapses to
\[
\psi_- = \Gh_0^\reg\left(\Gh_0^\sing\right)^{-1}\psi_+.
\]
In other words, the shift operator is then simply
\begin{equation}\label{magicidentity0}
\Shat = \Gh_0^\reg\left(\Gh_0^\sing\right)^{-1}
=\Gh_1^\reg\left(\Gh_1^\sing\right)^{-1}.
\end{equation}
In fact there is one further, very suggestive simplification 
in this case. 

Consider the exterior problem, in which  
boundary data $(\varphi(s),\nu(s)=\dydxh{\varphi}{n})$ are once again
supplied on $\partial\Omega$
but where the Helmholtz equation is now assumed to apply in the exterior 
region $\Omega'$, with radiating boundary conditions imposed 
at infinity. We change the notation for this boundary data to emphasise
that the solution of this exterior problem is different to that
of the interior one. Then, assuming that the normal is still defined to be 
pointing out of $\Omega$ (and into $\Omega'$) the boundary integral 
equation
\begin{equation}\label{ext}
\varphi = -\Gh_0\nu + (\Ihat+\Gh_1)\varphi
\end{equation}
replaces (\ref{bie}). Here $\Gh_1$ denotes
the same operator used elsewhere in this paper --- in which the 
limit $\x\to s$ in (\ref{defG1}) is taken from the 
\textit{interior}. The main differences from (\ref{bie}) are some sign 
changes to reflect that the normal points \textit{into} $\Omega'$
and the addition of an identity operator to the last term to 
correct the jump condition. Defining
$\Gh_{0,1}^\sing$ and $\Gh_{0,1}^\reg$ exactly as in the interior 
case, and letting the decomposition $\varphi=\varphi_-+\varphi_+$ be such that
\[
\varphi_+ = -\Gh_0^\sing\nu + (\Ihat+\Gh_1^\sing)\varphi,
\]
we get the analogue
\[
\left(\Ihat+\Gh_0^\reg\left(\Gh_0^\sing\right)^{-1}-\Gh_1^\reg+\Gh_0^\reg
\left(\Gh_0^\sing\right)^{-1}\Gh_1^\sing
\right)
\varphi_-
\]\begin{equation}\label{genscatext}\qquad\qquad\qquad\qquad
= \left(
\Gh_1^\reg
-\Gh_0^\reg\left(\Gh_0^\sing\right)^{-1}\Gh_1^\sing
\right)\varphi_+
\end{equation}
of (\ref{genscat}). In this case, if we assume that
$\Gh_0^\sing$ and $\Gh_1^\sing$ are chosen so that
(\ref{G1bis}) holds, we find that
\begin{equation} \label{ext-psi-}
\varphi_-\equiv 0.
\end{equation}
In other 
words, imposing (\ref{G1bis}) means that the regularised decomposition 
correctly identifies the component on the boundary that extends
to the outgoing component in the farfield.

\subsection{Further notation for the regularised decomposition}\label{uninot}
We now suggest notation that makes a 
more direct analogy with the notation already used for the 
one-dimensional problem and for the primitive decomposition. 
This notation also simpifies the representation of boundary conditions
as reflection operators in the model problems of
Secs.~\ref{egsec} and \ref{sec:dielcav}.

Let us 
define operators $\Dhat_0$ and $\Dhat_1$ by
\begin{eqnarray}
\Gh^\sing_0&=& \frac{\rmi}{2k}\Dhat_0 \label{D0}\\
\Gh^\sing_1&=& -\frac{1}{2}\left(\Ihat+\rmi\Dhat_1\right)
\qquad\left(\mbox{or}\quad
\Kh_1^\sing=-\frac{\rmi}{2}\Dhat_1\right). \label{D1}
\end{eqnarray}
That is, $\Dhat_0$ and $\Dhat_1$ are simply scaled and shifted
representations of the singular parts of the Green operators.
We further define operators $\Shat_0$ and $\Shat_1$ by
\begin{eqnarray*}
\Gh_0^\reg&=& \frac{\rmi}{2k}\Shat_0\Dhat_0\\
\Gh_1^\reg&=& -\frac{1}{2}\Shat_1\left(\Ihat+\rmi\Dhat_1\right).
\end{eqnarray*}
These equations replace (\ref{G01d}) and (\ref{G11d}) for the 
one-dimensional calculation and  (\ref{defT0}) and (\ref{defT1}) 
for the primitive decomposition. In particular, the primitive version is 
retrieved by replacing $\Dhat_0\to 1/\pnhat$ and $\Dhat_1\to 0$.
Note also that we can alternatively write
\begin{eqnarray*}
\Shat_0 &=& \Gh_0^\reg\left(\Gh_0^\sing\right)^{-1}\\
\Shat_1 &=& \Gh_1^\reg\left(\Gh_1^\sing\right)^{-1},
\end{eqnarray*}
which may be distinct if (\ref{G1bis}) is not imposed, as is the case for 
the primitive decomposition.

We confine our attention once again to the interior problem.
Then the outgoing and incoming components of the boundary solution are
respectively
\begin{equation}\label{defoutcirc}
\psi_+ = \frac{\rmi}{2k} \Dhat_0 \mu + \frac{1}{2}(1+\rmi\Dhat_1)\psi
\end{equation}
and
\[
\psi_- = \frac{\rmi}{2k} \Shat_0 \Dhat_0 \mu 
+ \frac{1}{2}\Shat_1(1+\rmi\Dhat_1)\psi.
\]
Eliminating $\psi$ and $\mu$ in favour of $\psi_-$ and $\psi_+$ in
these equations leads to (\ref{genscat}), in which the shift operator
takes the form
\[
\Shat=\left(\Ihat+\frac{1}{2}
(\Shat_0-\Shat_1)(\Ihat+\rmi\Dhat_1)\right)^{-1}
\frac{1}{2}\left(
\Shat_0(\Ihat-\rmi\Dhat_1)+\Shat_1(\Ihat+\rmi\Dhat_1)\right).
\]
This generalises the primitive equation (\ref{defT}).
If identity (\ref{G1bis}) is imposed then this reduces to the 
much simpler form
\begin{equation}\label{magicidentity}
\Shat = \Shat_0 = \Shat_1 
\end{equation}
previously provided in (\ref{magicidentity0}). These operators are
evaluated explicitly for the case of a circular domain in the next section.

\section{Applications to a model problem - circular cavities }\label{egsec}
In order to illustrate the application of the in-out decomposition to
a nontrivial problem, we consider now the case of the circle with
variable Robin boundary conditions 
of the form
\begin{equation}\label{robin}
\mu(s) = kF(s)\psi(s)\equiv k\Fhat\psi(s)
\end{equation}
with
\begin{equation}\label{coscase}
F(s) = a+b\cos\frac{s}{ R}.
\end{equation}
This includes as a special case $a=0$
the examples treated in \cite{BerryDennis,Dubphase}. Such problems
provide an interesting challenge for our approach when regions of 
the boundary exist for which $F(s)>0$. This is because the system
then supports modes that are localised in the full 2D problem near the 
boundary \cite{BerryDennis,Dubphase} and that have on the boundary
itself significant
components in momentum representation with $p^2>1$. The regions around 
and beyond critical lines $p^2=1$ then play a significant role and allow us to 
test decompositions across
the transition from open to closed modes.
We emphasise that, although the Green function decomposition can be achieved 
analytically in this geometry, the variable boundary conditions make 
the complete problem nonintegrable \cite{Dubphase}. 
\subsection{Shift operator for the circle}\label{circsec}
We begin by finding the shift operator for the circle.
In this case we can write explicit analytical results for
the various operators defined in Secs.~\ref{unishift} and \ref{uninot}. Furthermore,
we find that identity (\ref{G1bis}) automatically holds when 
$\Gh_0^\sing$ and $\Gh_1^\sing$ are respectively defined by
(\ref{singcont}) and (\ref{G1full}).

The chord-length function for a circle of radius $R$ is
\[
L(s,s') = 2R\sin\frac{|s-s'|}{2R}.
\]
Then an application of Graf's theorem \cite{AS} allows us to write the 
kernel of (\ref{defG0}) in the form
\begin{eqnarray*}
\lim_{\x\to s}G_0(\x,s';k) 
&=& \frac{\rmi}{4}\sum_{m=-\infty}^\infty H_m(z)J_m(z)\e^{\rmi m(s-s')/R},
\end{eqnarray*}
where $z=kR$. In a Fourier basis (or momentum representation), $\Gh_0$
is then represented by a diagonal matrix whose entries are
\begin{eqnarray*}
(\Gh_0)_{mm} &=& \frac{\rmi\pi z}{2k}H_m(z)J_m(z) =\frac{\rmi\pi z}{4k}
\left(|H_m(z)|^2+H_m(z)^2\right).
\end{eqnarray*}
The two alternative forms given here for $(\Gh_0)_{mm}$ 
respectively provide the decomposition of $\Gh_0$ for $|m|>z$ 
and for $|m|<z$. That is, one finds (although we do not show the detailed 
integrations here) following the recipe described in 
Sec.~\ref{contoursec} that the singular and regular parts of the Green
operator $\Gh_0$ are represented by diagonal matrices with entries
\begin{eqnarray*}
(\Gh_0^\sing)_{mm} &=& \frac{\rmi}{2k}
\cases{\frac{\pi z}{2}|H_m(z)|^2
& if $|m|\leq z$\\
{\pi z}H_m(z)J_m(z)
&if $|m|> z$}
\end{eqnarray*}
and 
\begin{eqnarray*}
(\Gh_0^\reg)_{mm} &=& \frac{\rmi}{2k}
\cases{\frac{\pi z}{2}H_m(z)^2
& if $|m|\leq z$\\ 
0&if $|m|> z$.}
\end{eqnarray*} 
Similarly, one finds that
\[
(\Gh_1)_{mm} = \frac{\rmi\pi z}{2}H_m'(z)J_m(z)
\] 
has the decomposition
\begin{eqnarray*}
(\Gh_1^\sing)_{mm} &=& -\frac{1}{2}
\cases{\frac{\pi z}{2\rmi}H_m'(z)H_m^*(z)
& if $|m|\leq z$\\
\frac{\pi z}{\rmi}H_m'(z)J_m(z)
&if $|m|> z$}
\end{eqnarray*}
and
\begin{eqnarray*} 
(\Gh_1^\reg)_{mm} &=& -\frac{1}{2}
\cases{\frac{\pi z}{2\rmi}H_m'(z)H_m(z)
& if $|m|\leq z$\\
0&if $|m|> z$.}
\end{eqnarray*}
We also find that
\begin{eqnarray}
\left(\Gh_0^\reg\left(\Gh_0^\sing\right)^{-1}\right)_{mm}&=&
\left(\Gh_1^\reg\left(\Gh_1^\sing\right)^{-1}\right)_{mm}\nonumber \\
&=&\cases{\frac{H_m(z)}{H_m^*(z)} &if $|m|\leq z$\\
0&if $|m|> z$.} \label{S-circ}
\end{eqnarray}
In summary, we have shown that
\[
(S)_{mm} 
= (S_0)_{mm}= (S_1)_{mm}
\]
holds in the circular case (as does (\ref{G1bis})).

It is also useful to record that the scaled singular parts 
defined in Sec.~\ref{uninot} have diagonal elements
\[
\left(D_0\right)_{mm} 
 =\frac{\pi z}{2} f_m(z)
\]
and
\[
\left(D_1\right)_{mm} =-\frac{\pi z}{4} f_m'(z),
\]
where
\[
f_m(z) = \cases{|H_m(z)|^2
&if $|m|\leq z$\\
2H_m(z)J_m(z)&if $|m|> z$}
\]
and primes indicate derivatives with respect to $z$.
We note that the leading Debye approximation for 
$H_m(z)$ \cite{AS} yields, for $z\gg 1$, 
\[
(D_0)_{mm} \simeq \left(\pnhat^{-1}\right)_{mm} =
\cases{\frac{1}{\sqrt{1-m^2/z^2}}&if $|m|< z$\\
\frac{1}{\rmi\sqrt{m^2/z^2-1}} &if $|m|> z$},
\]
consistent with the primitive decomposition on making the
identification $p=m/z$, while
\[
(D_1)_{mm} \simeq \frac{1}{2z} D_0^3
\]
is asymptotically of higher order as $z=kR\to\infty$. Such primitively
defined operators have been successfully used in \cite{Hanyathesis} to 
describe eigenfunctions of a disk with boundary conditions changing 
discontinuously from Dirichlet to Neumann, where diffractive effects
are important. We show in the more detailed calculation of
Sec.~\ref{robsec}, however, that they lead to an in-out decomposition
with undesirable features in the evanescent region of momentum space
(see Fig.~\ref{psiprimfigs}).

\subsection{Circular cavity with variable Robin boundary conditions}\label{robsec}

Having characterised the shift operator in the previous section, 
the remaining step is to formulate the prescribed boundary 
conditions in terms of a reflection operator mapping the incoming 
to the outgoing boundary solution. To this end we simply
substitute the boundary condition (\ref{robin}) into the equation
(\ref{defoutcirc}) that defines the outgoing wave $\psi_+$, which
leads to
\[
\left(\Ihat-\rmi\Dhat_1-\rmi\Dhat_0\Fhat\right)\psi_+
= \left(\Ihat+\rmi\Dhat_1+\rmi\Dhat_0\Fhat\right)\psi_-.
\]
Hence
\begin{equation}\label{refl}
\psi_+=\left(\frac{\Ihat+\rmi\Xhat}{\Ihat-\rmi\Xhat}\right)\psi_-
\equiv\rh\psi_-,
\end{equation}
where
\[
\Xhat = \Dhat_0\Fhat+\Dhat_1.
\]
Note that $\Xhat$ is similar to the operator 
\[
\Xhat' = \Dhat_0^{1/2}\Fhat\Dhat_0^{1/2}+\Dhat_1
\]
whose projection onto the open subspace $\H_\o$ is Hermitian, so that
the corresponding eigenvalues of $\rh$ lie on the unit circle.
Eigenmodes are now obtained as solutions of $\psi_+=\rh\Shat\psi_+$.
The secular equation may in this case be more conveniently written
\[
\det\left(1-\rmi\Xhat-(1+\rmi\Xhat)\Shat\right)=0;
\]
the evaluation is straightforward here, requiring  
simply the determinant of a tridiagonal matrix for the special case
(\ref{coscase}). Alternatively, the secular equation can be written directly
as a difference equation as described in \cite{BerryDennis,Dubphase},
albeit expressed in a different basis to that used here. We note that 
the form of the reflection operator in (\ref{refl}), is independent of the
 choice of $\Fhat$ in  (\ref{robin}). Our decomposition
 in terms of shift and reflection operators thus allows us to treat general 
 boundary conditions easily once the shift operator has been
 constructed. (Of course, the transfer operator does not have tridiagonal 
 form in general.)

\subsection{Evanescence in solutions}\label{evansec}
We now describe some particular solutions of this model problem,
concentrating on their behaviour in the region of closed modes
$p^2>1$.

\begin{figure}[h]\begin{center}
\includegraphics[scale=0.6]{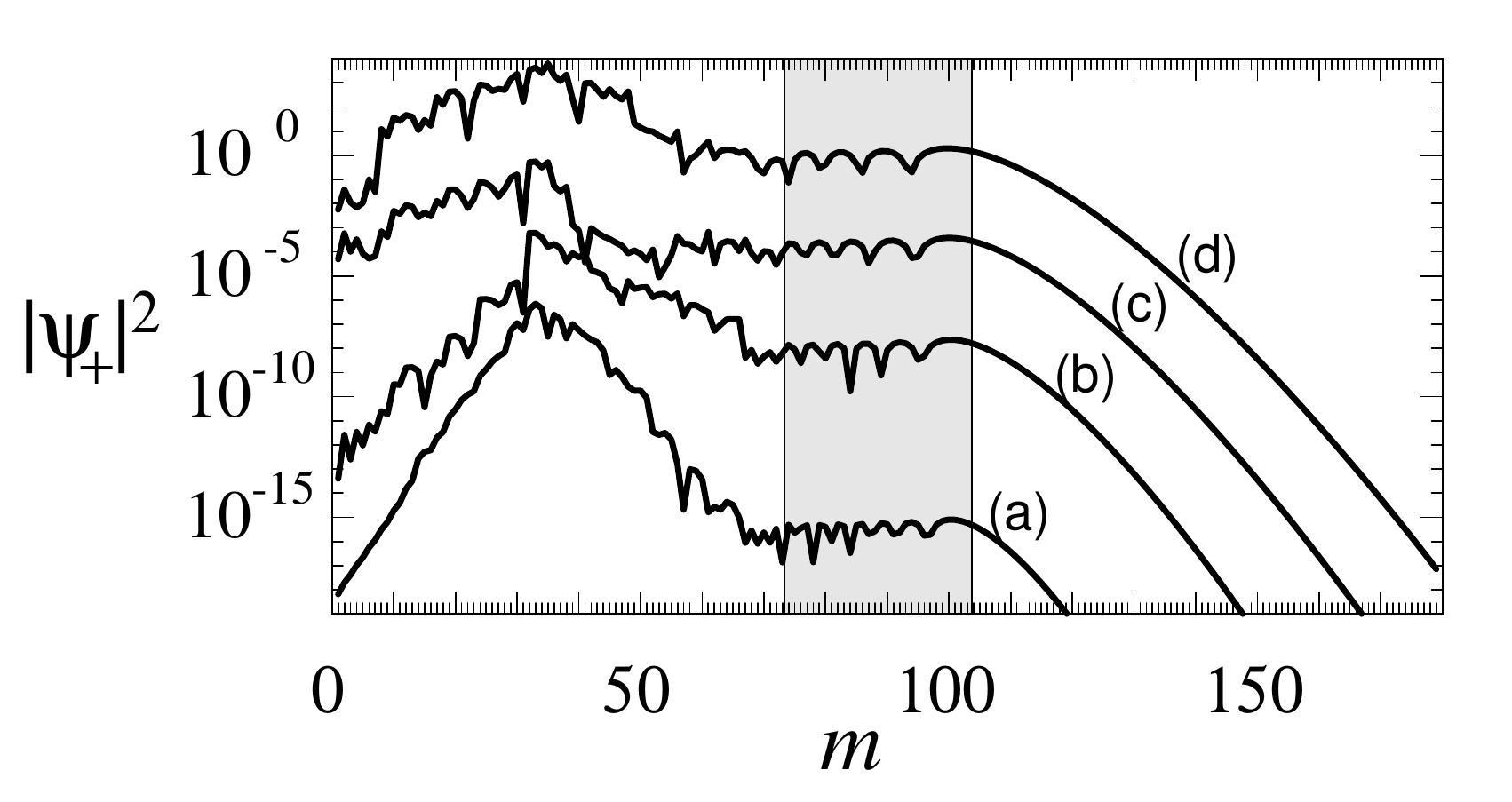}
\end{center}
\caption{Outgoing boundary intensities are shown in momentum representation for
sample eigenfunctions with $(a,b)=(0,1)$ in (a), $(a,b) =(-1,2)$ in (b),
$(a,b) =(-2,3)$ in (c) and $(a,b) =(-3,4)$ in (d). The respective
wavenumbers 
are $k\approx 73.32$ in (a), $k\approx 73.84$ in (b), 
$k\approx 74.06$ in (c) and $k\approx 74.17$ in (d). 
 Successive examples have been scaled by
arbitrary factors to separate them graphically. All of these examples
have $F_\min=0$, $F_\max=1$ and similar values of $kR\approx 74$ and therefore share a
common  ``allowed region'' associated with the boundary-localised
waves (\ref{boundarywave}), occupying the band $kR<m<\sqrt{2} kR$ and  indicated by the
shaded region in the figure.
}\label{psifigs}
\end{figure}

Figure~\ref{psifigs} shows four examples of $\psi_+$-eigenfunctions for
boundary condition (\ref{coscase}), corresponding respectively to 
$(a,b)=(-3,4)$, $(-2,3)$ $(-1,2)$ and $(0,1)$.
A prominent feature of all four examples is that, 
following decay from a peak around $m=33$ to the
border at $m=kR\approx 74$ between open and closed modes, there follows a
plateau extending significantly into the closed region $m>kR$. 
This feature is explained in \cite{BerryDennis,Dubphase} in terms of
waves localised near the boundary wherever $F(s)>0$. We now
offer a brief summary of this explanation, borrowing from the
discussions in \cite{BerryDennis,Dubphase}.

Consider first the case where $F>0$ is a constant and $kR\gg 1$. Then
one can find quasimodes decaying exponentially into the interior
(denoting $p=m/kR$ and $s=R\theta$) of the form
\begin{equation}\label{boundarywave}
\psi(r,\theta) \simeq \e^{-kq(R-r)+\rmi m\theta}=\e^{-kq(R-r)+\rmi kps},
\end{equation}
satisfying the interior wave equation if $p^2=1+q^2$ and 
the boundary conditions if $q=F$. Note that such boundary-localised
solutions occupy the closed subspace corresponding to $p^2>1$. If $F$ now
varies, but over a scale longer than that of a wavelength, we should therefore
expect localised solutions formed by deformations of these states, spread in
momentum representation over the region
\[
1+F_\min^2<p^2<1+F_\max^2,
\]
 where 
\begin{eqnarray*}
F_\max &=& \max(\max(F(s)),0)\\
F_\min &=& \max(\min(F(s)),0).
\end{eqnarray*}
Exact eigenfunctions will then generically couple to such local solutions
and lead to the plateau structures shown in
Fig.~\ref{psifigs}. 
Heuristically we can interpret the observed 
plateaux as being the result of tunnelling in momentum space from the 
primary quasimode around $m=33$ to an ``allowed region''  
$1+F_\min^2<p^2<1+F_\max^2$ introduced by the boundary-localised mode. 
In fact, the four examples in Fig.~\ref{psifigs} were chosen to have the same
values of $F_\min=0$ and $F_\max=1$, and one does indeed then observe that the plateaux occupy
the same interval, indicated by the shaded strip in the
figure. 

\begin{figure}[h]\begin{center}
\includegraphics[scale=0.6]{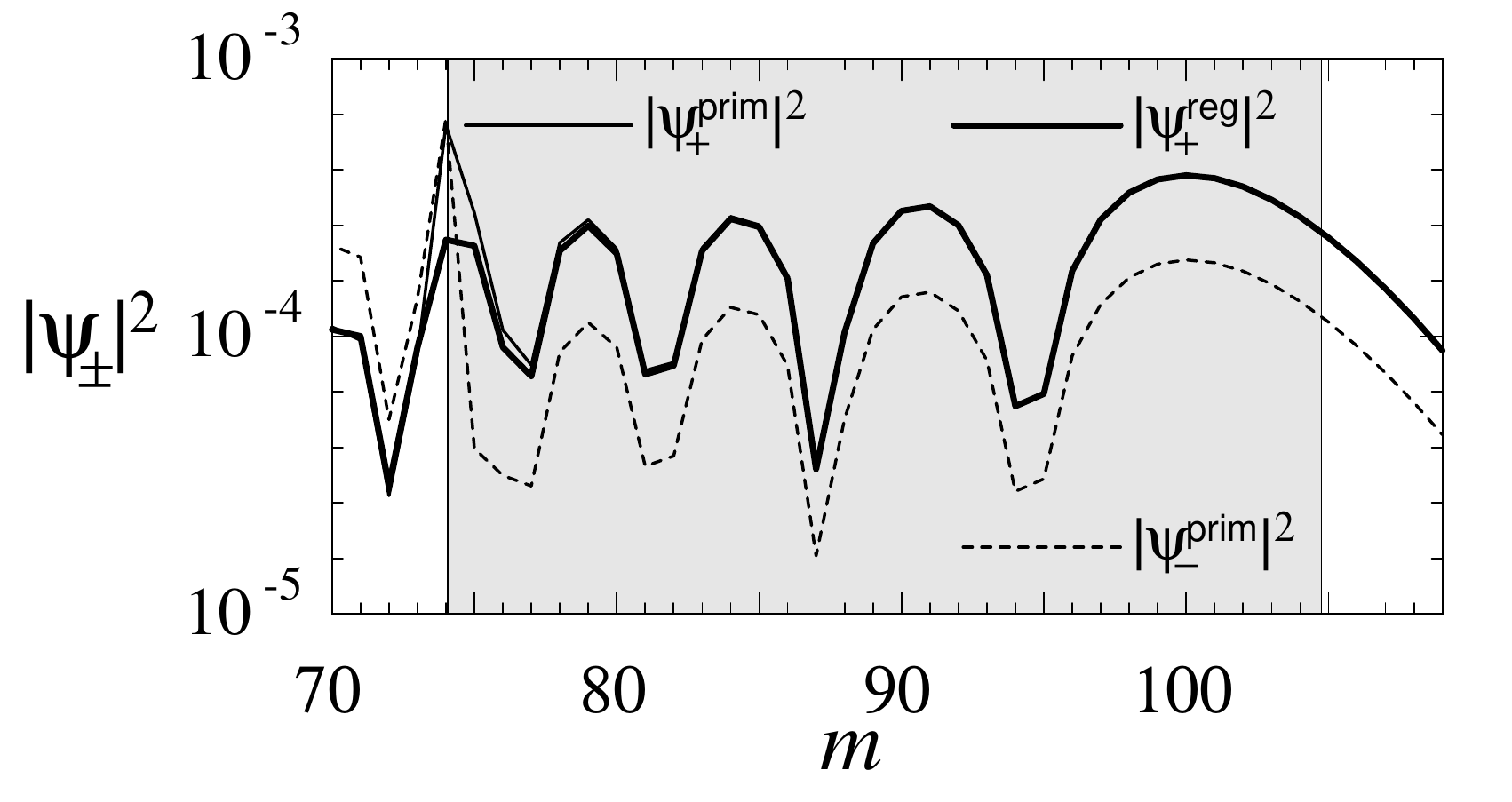}
\end{center}
\caption{The outgoing boundary intensity $|\psi_+^\reg|^2$ defined by the
regularised decomposition is shown as a heavy continuous curve over
the plateau region for case (c) of Fig.~\ref{psifigs}:
as in
Fig.~\ref{psifigs}, the ``allowed region'' $1<p^2<F_\max^2$ is 
indicated by the shaded strip. 
Also shown, as a thin continuous curve, is the version
$|\psi_+^\prim|^2$ 
obtained from the primitive 
decomposition. The thin dashed curve shows the primitive \textit{incoming} 
intensity $|\psi_-^\prim|^2$. The primitive and regularised  outgoing intensities differ 
significantly only near $m= kR\approx 74$, where the primitive case shows an
expected divergence. The smaller, but significant, incoming intensity
observed in the primitive case is qualitatively different from the 
regularised case, where the incoming intensity is zero or exponentially
smaller than the outgoing intensity (depending on the conventions used
to define the contour $\Gamma_0$ in Sec.~\ref{contoursec}).
}\label{psiprimfigs}
\end{figure}

This example offers a useful illustration of the most significant
differences between the regularised and primitive decompositions 
--- and their corresponding definitions of $\psi_\pm(s)$. In
Fig.~\ref{psiprimfigs} we have selected the eigenfunction corresponding
to case (c) of Fig.~\ref{psifigs} and shown it  (as the thick
continuous curve) in more detail over its plateau region 
$1<p^2<F_\max^2$. Also shown are the primitive version (as a thin
continuous curve) and the primitive incoming solution (as the thin
dashed curve). As expected, the regularised and primitive outgoing intensities are
very close, hardly distinguishable on this log plot, except 
that the primitive intensity becomes significantly larger in the
region of critical incidence $p^2\approx 1$. We emphasise that this divergence is an
artefact of the primitive decomposition, and its failure to deal properly
with curvature at critical reflection, rather than an intrinsic
characteristic of the solution itself. 

The second difference is that,
in the case of primitive decomposition, we obtain a smaller, but 
nonetheless significant, incoming component even in the ``evanescent'' 
region $p^2>1$, see Fig.~\ref{psiprimfigs}. By contrast, the incoming component vanishes for the 
regularised decomposition. 
The nonvanishing
primitively-defined incoming wave can be understood further by
explicitly evaluating the shift operator for the primitive
decomposition. Evaluating expression (\ref{defT}) for the special case of
the circle yields a primitive shift operator represented by a diagonal
matrix with entries
\begin{equation}\label{Sprim}
(\Shat^\prim)_{mm} = 
\frac{\sqrt{1-m^2/z^2}J_m(z)-\rmi J_m'(z)}{\sqrt{1-m^2/z^2}J_m(z)+\rmi J_m'(z)};
\end{equation}
which needs to be compared with $\Shat$ in (\ref{S-circ}) for the regularised 
decomposition. 
For closed components with $m^2>z^2\equiv(kR)^2$, and using the Debye
approximation for the numerator, Eq.\ (\ref{Sprim}) vanishes at leading order. However,
the next-to-leading order contributions do not cancel. Therefore
$\psi_-^\prim=\Shat^\prim\psi_+^\prim$  does not vanish as does the
regularly-defined incoming wave, or even decay exponentially as 
one would expect for properly-defined evanescent waves. Instead, it
is only smaller than $\psi_+^\prim$ by a factor scaling algebraically with $1/k$.
This failure of the primitive
decomposition to separate evanescently incoming and outgoing solutions
at all orders puts it at a significant disadvantage at describing
such solutions compared to the regularised decomposition.

\subsection{Husimi functions and Goos-H\"anchen shifts}

A particularly helpful means of presenting eigenmodes of 
this problem is by the boundary Husimi functions. Here we are 
guided by previous applications to dielectric problems
\cite{NoeckelStone,BFSHus,Hus,WHforHus}, where a definition of boundary
Husimi functions has been proposed. This amounts in our 
language to applying the conventional definition of Husimi 
functions to the primitive in-out components 
\begin{equation}\label{defPsiprim}
\Psi^\prim_\pm=\sqrt{\pnhat}\psi_\pm
\end{equation} 
discussed further in \ref{Appsc}. In this paper we choose to work
instead 
with a variation 
\begin{equation}\label{defPsi}
\Psi_\pm=\Dhat_0^{-1/2}\psi_\pm
\end{equation} 
in
which $\pnhat^{1/2}$ is replaced by its regularised analogue 
$\Dhat_0^{-1/2}$. That is, we define
\[
H_\pm(s_0,p_0) = |\braket{s_0,p_0}{\Psi_\pm}|^2,
\]
where the state $\ket{s_0,p_0}$ corresponds to the boundary function
\[
\braket{s}{s_0,p_0} = \sum_{n=-\infty}^\infty
\left(\frac{k}{2\pi b^2}\right)^{1/4}\e^{-k(s-s_0-n\ell)^2/4b^2+\rmi kp_0(s-s_0-n\ell)}
\]
and the inner product is 
\[
\braket{\Psi}{\Phi} = \int_0^\ell \Psi^*(s)\Phi(s)\d s.
\]
Using (\ref{defPsi}) rather  than (\ref{defPsiprim}) gives a Husimi function that is
better behaved around the critical line $p^2=1$ but is otherwise very
similar in the region $p^2<1$ and does not have a significant impact
on the discussion following.

\begin{figure}[h]\begin{center}
\includegraphics[scale=0.7]{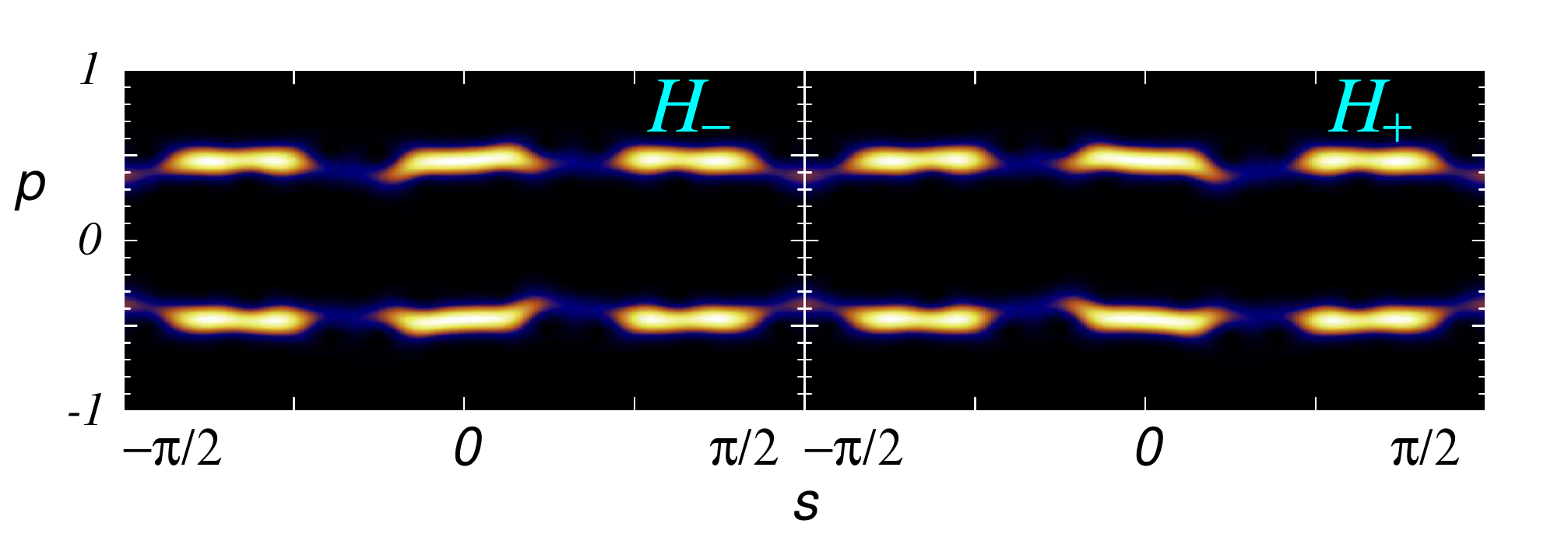}
\end{center}
\caption{Boundary Husimi representations are shown of the outgoing and incoming
solutions for the example labelled (d) in Fig.~\ref{psifigs}.}\label{husallfig}
\end{figure}

A particular pair of Husimi functions $H_+(s,p)$ and $H_-(s,p)$
calculated in this way is
shown in Fig.~\ref{husallfig}, corresponding to the eigenmode labelled
(d) in Fig.~\ref{psifigs}. The incoming and outgoing modes in the
figure are qualitatively similar but display differences in detail that 
are explained by the application of the reflection operator
in (\ref{refl}) --- or equivalently by the Goos-H\"anchen (GH) shift 
\cite{SH,elloops,nonHam,NonHam2,GFIEEE,GFQE}. Goos-H\"anchen and
related perturbations of ray dynamics are typically explained in terms
of a shifting of the centre of a Gaussian or other localised beam following
reflection from a boundary or interface. Such approaches provide a
clear physical interpretation of the effect but seem unnatural in the
context of calculating eigenfunctions, where no Gaussian beams appear
in the rather more complicated wave patterns that one finds in typical 
cases. We now point out that the in-out decomposition of boundary
solutions explored in this paper provide a much more direct way of
describing the effect in this context. It makes it possible to explain
the GH shift
in terms of the tools that are naturally used to calculate
eigenfunctions with chaotic or otherwise nonintegrable ray dynamics.
Furthermore, the improved treatment of critical reflection such as
provided by the regularised decomposition may allow improved
descriptions of effects such as Fresnel filtering that are observed
in associated parts of phase space, although we do not pursue that
issue explicitly in this paper.

The incoming and outgoing  solutions respectively satisfy the
equations
\[
\rh\Shat \psi_+ = \psi_+
\]
and 
\[
\Shat\rh\psi_- = \psi_-,
\]
where $\Shat$ describes the ray dynamics of specular reflection and is
common to all cavity problems occupying the same domain, irrespective
of boundary conditions. The GH shift may be incorporated
simply by representing $\rh$ in the exponential form (\ref{reflgen}) and regarding
it as a quantisation of the flow of a Hamiltonian generator $h(s,p)$.
The complete dynamics appropriate to $\psi_\pm$ are then obtained simply as
a stroboscopic intertwining of this ``Goos-H\"anchen map'' and the regular,
specular ray dynamics --- albeit in opposite orders for $\psi_+$ and $\psi-$.

The reflection operator in (\ref{refl}) is generated by
\[
\hh = \frac{\rmi}{k} \log\frac{1+\rmi\Xhat}{1-\rmi\Xhat}.
\]
Away from the critical line $p^2=1$ this has the leading-order
``primitive'' symbol
\[
h(s,p)  = 
\frac{\rmi}{k} \log\frac{1+\rmi F(s)/\sqrt{1-p^2}}{1-\rmi F(s)/\sqrt{1-p^2}}
 = \frac{\rmi}{k} \log\frac{\sqrt{1-p^2}+\rmi F(s)}{\sqrt{1-p^2}-\rmi F(s)}
\]
and the Goos-H\"anchen map can be approximated at leading order by
\begin{eqnarray}\label{GHvec}
\Delta s &\approx& \dydxv{h}{p} = -\frac{2}{k}\frac{F(s)p/\sqrt{1-p^2}}{1-p^2+F(s)^2}\nonumber\\
\Delta p &\approx& -\dydxv{h}{s} = -\frac{2}{k}\frac{F'(s)\sqrt{1-p^2}}{1-p^2+F(s)^2}.
\end{eqnarray}
Note that $\Delta s$ and $\Delta p$ are $O(1/k)$ and therefore act as
perturbations of the usual specular ray dynamics.  Note also that there is a
nontrivial shift in momentum as well as position here because the
local reflection phase depends on $s$. In the case of planar
reflection, where the reflection phase is constant along the
interface, the Goos-H\"anchen map changes $s$ but not $p$, as
prescribed in classical descriptions of the effect.

\begin{figure}[h]\begin{center}
\includegraphics[scale=0.7]{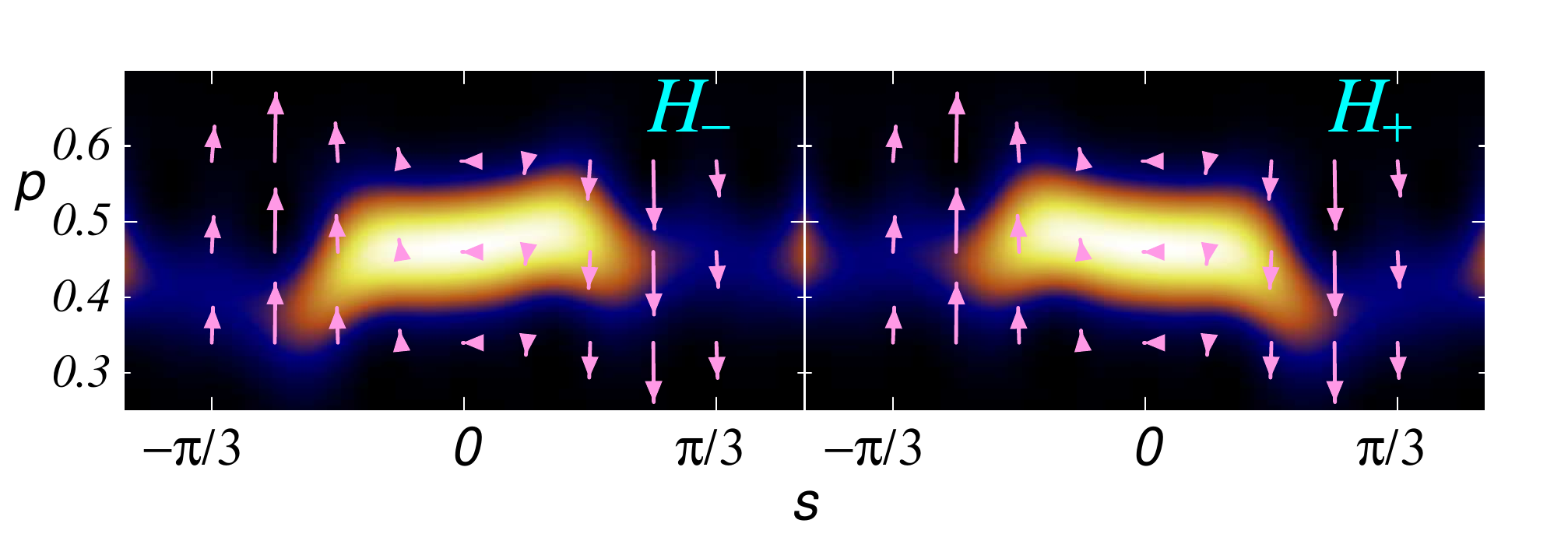}
\end{center}
\caption{An expanded view is given of the middle-upper lobes of
  Fig.~\ref{husallfig}. Superimposed is the Hamiltonian vector field of
  the Goos-H\"anchen map, defined explicitly by (\ref{GHvec}).}\label{husvecfig}
\end{figure}

The breaking by the boundary conditions of integrability is
strong enough in the example of Fig.~\ref{husallfig} 
 to localise the mode within an island chain of 
Goos-H\"anchen-perturbed ray dynamics (see \cite{elloops} for a
treatment of this 
effect in the context of dielectric cavities). 
In Fig.~\ref{husvecfig} we illustrate the Goos-H\"anchen vector field
$(\Delta s,\Delta p)$ over part of phase space surrounding the point
$(s,p)=(0,1/2)$, which lies at the centre of the middle-upper lobes of the Husimi
functions of Fig.~\ref{husallfig}: this is where the incoming
and outgoing Husimi functions differ most in this example. One sees
that the vector field is indeed entirely consistent with the
deformation of the incoming Husimi function relative to its outgoing equivalent:
features lying at the bases of the Hamiltonian vectors for the former case lie at
their tips in the latter case. Note that Poincar\'e plots of the 
corresponding dynamics (Goos-H\"anchen map following the specular map
in the outgoing case and preceding it in the incoming case) show a
similar deformation (not shown).

\section{Application to dielectric cavities}
\label{sec:dielcav}

Our final application is to the problem of calculating resonant modes in
dielectric cavities. The problem is solved explicitly here for the
circular cavity with boundary conditions appropriate to TM
polarisation. This simple problem can be solved analytically
by elementary means: our aim, however, is to illustrate the application of
the method to inside/outside problems in a way that it might later be
generalised to less trivial, noncircular domains.

In this case, we must separately solve two wave problems; the interior
one whose solution and normal derivative we denote as usual by
$\psi$ and $\mu$ respectively, and which is such that
\[
-\nabla^2\psi = n^2 k^2\psi, \qquad\x\in\Omega,
\] 
and the exterior problem whose solution
and normal derivative are respectively denoted by $\varphi$ and $\nu$
and for which
\[
-\nabla^2\varphi = k^2\varphi,\qquad\quad  \x\in\Omega'.
\]
Here the refractive index inside the cavity is $n>1$ while outside we
set $n=1$.
Following the discussion of Sec.~\ref{unishift}, the interior and
exterior problems separately define decompositions
$\psi=\psi_-+\psi_+$ and $\varphi=\varphi_-+\varphi_-$ and shift operators
such that
\[
\psi_- = \Shat^\inner\psi_+
\]
and
\[
\varphi_- = \Shat^\ext\varphi_+.
\]
The shift operators $\Shat^\inner$ and $\Shat^\ext$ differ by using
Green functions corresponding to different values of the refractive
index and by the boundary integral equation for the exterior problem
having a unit normal pointing into rather than out of the domain 
concerned, (see Eqn.\ ({\ref{ext})). According to the conventions
suggested in Sec.~\ref{unishift}, where the endpoint contour $\Gamma_0$ 
is chosen for closed modes to account for all of the boundary integral, we find
that the external shift operator vanishes,
\[
\Shat^\ext = 0
\]
and Eqn.\ (\ref{ext-psi-}) holds (see \ref{Appalt} for discussion of
alternative conventions). Of course it is the field $\varphi(\x)$ in
the exterior $\x\in\Omega'$ that is physically observed here: see 
\ref{Appint} for a discussion of the determination of the full
solution, inside and outside the cavity, from its restriction to the boundary.

The end result will be presented more compactly if we use the shift 
operator for the external problem to
define a Dirichlet-to-Neumann map $\Fhat$, such that
\begin{equation}\label{D2N}
\nu = k \Fhat\varphi.
\end{equation}
In general (after some manipulation),
\begin{equation}\label{defF}
\Fhat = \rmi\left(\Dhat_0^\ext\right)^{-1}\left(\Ihat+\rmi\Dhat_1^\ext\right).
\end{equation}
and we note that $\Fhat$ itself is independent of conventions
used in the in-out decomposition. Note also that $\Fhat$ is determined
by the condition of having no incoming waves at infinity and is
independent of boundary conditions on the cavity itself.

We next apply the cavity's boundary
conditions. To be concrete, we impose TM boundary conditions 
\[
\psi(s) = \varphi(s)\qquad\mbox{and}\qquad\mu(s) = \nu(s),
\]
but emphasise that the following discussion generalises
easily. Together with (\ref{D2N}), this provides a boundary condition
formally very much like that of Sec.~\ref{egsec} and leads to a similar-looking
reflection operator
\[
\rh = \frac{1+\rmi\Xhat}{1-\rmi\Xhat},
\]
except that, in this case,
\[
\Xhat = \Dhat^\inner_1+\frac{1}{n}\Dhat^\inner_0\Fhat.
\]
Note that this reflection operator can alternatively be written
\begin{equation}\label{defrh}
\rh = \left(n(\Dhat_0^\inner)^{-1}(\Ihat-\rmi\Dhat_1^\inner)-\rmi\Fhat\right)^{-1}
\left(n(\Dhat_0^\inner)^{-1}(\Ihat+\rmi\Dhat_1^\inner)+\rmi\Fhat\right).
\end{equation}
This allows for a more direct comparison with classical
expressions for the corresponding Fresnel reflection coefficient
\[
r = \frac{n\cos\alpha-\cos\beta}{n\cos\alpha+\cos\beta}.
\]
This comparison is achieved on making the identifications
\[
(\Dhat_0^\inner)^{-1}(\Ihat-\rmi\Dhat_1^\inner)
\approx
(\Dhat_0^\inner)^{-1}(\Ihat+\rmi\Dhat_1^\inner)
\sim\cos\alpha\equiv\sqrt{1-p^2}
\] 
and
\[
-\rmi\Fhat=(\Dhat_0^\ext)^{-1}(\Ihat+\rmi\Dhat_1^\ext)
\sim\cos\beta\equiv\rmi\sqrt{n^2p^2-1}
\]
appropriate to the ray-dynamical limit, where $\alpha$ and $\beta$
are respectively the angles of incidence and refraction of a ray
hitting the boundary from inside.

We can now determine resonant modes as solutions of the secular 
condition
\begin{equation}\label{wavemap}
\psi_+ = \rh\Shat^\inner\psi_+.
\end{equation}
This is once again formally much as in the model example of
Sec.~\ref{egsec}. 
The main qualitative difference is that, whereas the
restriction of the reflection operator to the open subspace has
eigenvalues on the unit circle in the example of Sec.~\ref{egsec},
here  the relevant eigenvalues of $\rh$ lie inside the unit circle.
reflecting the lossiness of the system (to radiation to infinity).
As a consequence, the solutions of the secular equation are obtained
here for complex values of $k$, as should be expected for resonant
modes with a finite lifetime.

We point out that composing the shift operator
$\Shat$ with a subunitary reflection operator $\rh$ as in (\ref{wavemap}) provides a very
natural wave analogue of the ray-dynamical approach taken, for
example, in \cite{SYscar0,SYscar,SYsurv,Shinohara1}. There, emission from dielectric cavities is
treated by iterating initial densities of rays under a composition
of the ray-dynamical analogues of $\Shat$ and $\rh$, namely a
Perron-Frobenius operator encoding the 
Poincar\'e-Birkhoff map of the internal dynamics and a transmission
loss determined from classical Fresnel
coefficients. The combined operator in (\ref{wavemap}) simply replaces
each of those elements by a corresponding wave map. It therefore
offers
a basis for treating inherently wave effects within the broader
approach of \cite{SYscar0,SYscar,SYsurv,Shinohara1}, particularly, for example, the 
more complicated transmission losses encountered around the critical
lines (see also further discussion below).

We now specialise these results to the case of a circular cavity. One
then finds from (\ref{defF}) that $\Fhat$ is diagonal and has elements
\[
(\Fhat)_{mm} = \frac{H_m'(kR)}{H_m(kR)}.
\]
(Of course this can also be deduced directly from (\ref{D2N}) by
elementary means but our aim here is to offer a discussion that 
generalises.) We also find that
\[
\left((\Dhat_0^\inner)^{-1}(\Ihat+\rmi\Dhat_1^\inner)\right)_{mm} = 
-\rmi\frac{ H_m'(nkR)}{H_m(nkR)}
\] 
(for all $m$) and
\[
\left((\Dhat_0^\inner)^{-1}(\Ihat-\rmi\Dhat_1^\inner)\right)_{mm} = 
\cases{\rmi\frac{ {H^*_m}'(nkR)}{H^*_m(nkR)},
& if $|m|\leq nkR$\\
\rmi\frac{ J_m'(nkR)}{J_m(nkR)},
&if $|m|> nkR$,}
\]
from which the reflection operator follows by (\ref{defrh}). Note that
this provides a means of calculating curvature corrections to Fresnel
laws, such as
described in \cite{HScirc}, while also incorporating the phase information
relevant to, for example,  Goos-H\"anchen shifts.

The elements $(\Fhat)_{mm}$ undergo a transition across the critical
lines of the exterior problem, defined by $|m|=kR$. They are
approximately imaginary for $|m|<kR$ and define through
(\ref{defrh}) diagonal elements of the reflection operator such 
that $|(\rh)_{mm}|<1$: corresponding resonant modes leak
directly by refraction to the exterior and are short-lived. Whispering
gallery modes lie within the open subspace of the interior problem,
for which $|m|<nkR$, and the closed subspace of the exterior problem,
for which $|m|>kR$. The corresponding elements 
\[
(\Fhat)_{mm} = \frac{Y_m'(kR)-\rmi J_m'(kR)}{Y_m(kR)-\rmi J_m(kR)}
\approx  \frac{Y_m'(kR)}{Y_m(kR)} +\rmi \frac{2}{\pi kR}\frac{1}{Y_m(kR)^2}
\]
of $\Fhat$ are
approximately real but retain small positive imaginary parts and lead to
diagonal elements of $\rh$ which are close to, but slightly less than,
unity in magnitude: the corresponding whispering gallery modes
therefore have a long but finite lifetime.

A representation of $\rh$ in exponential form (\ref{reflgen}) defines
 a generator $\hh$ with significant nonHermitian components
in the region $|m|<kR$ and smaller but nonvanishing nonHermitian 
components over $kR<|m|<nkR$, reflecting the lossiness of the
system as a whole. Away from the critical lines $|m|=kR$, where
the reflection phase is not rapidly varying, we may approximate this 
operator straightforwardly by a Goos-H\"anchen map for which the 
Hamiltonian generator $h(s,p)$ has imaginary components and is
obtained as the ray-dynamical symbol of $\hh=\rmi\log(\rh)/k$. 
Across the critical lines $|m|=kR$, where the reflection phase 
is rapidly varying, the ray-dynamical shifts corresponding to $\rh$ are 
less easy to work out. However, one can still formally use $\rh$
to define a ray-dynamical map --- it is just not as close to the 
identity map and its generator is not as easily approximated. Further
investigation of this issue will hopefully lead to a better understanding
of the modifications necessary to  incorporate by Goos-H\"anchen and 
related effects into a ray dynamics around the physically important 
regime of critical refraction.

In summary, we have shown how the in-out decompositions discussed in
this paper lead naturally to expected results for resonant modes of 
a dielectric circle. We emphasise, however, that the general approach
extends to deformed cavities. In particular, we demonstrate that this 
approach shows promise for a better understanding of
effects such as Goos-H\"anchen and similarly modified dynamics near 
critical lines; in the interests of brevity
we do not pursue this issue any further in this paper.

\section{Conclusions}\label{concsec}

We have proposed an approach for separating the boundary solutions of
cavity problems  into incoming and outgoing components that allows 
boundary integral equations to be recast as transfer-operator
equations. The resulting ``shift operator'' maps the outgoing component
 leaving the boundary onto the incoming component arriving back at
the boundary having crossed the cavity's interior. The shift operator is
completely independent of the boundary conditions and decouples from them
the problem of wave propagation through the interior. The
boundary conditions are used separately to complete the problem by defining a
reflection operator mapping the component arriving at the boundary to
the reflected, outgoing component leaving it again towards the interior.

The underlying intuition behind this approach is familiar in the
context of semiclassical approximations 
\cite{Bog}. We believe, however, that using (\ref{singeq}) and 
(\ref{regeq}) rather than ray-based ideas in order to define the 
in-out decomposition of the boundary solution provides an ideal 
platform on which to extend these ideas beyond leading-order 
approximations. We hope that the exactly defined shift operator will enable the
efficient treatment of  diffraction and tunnelling effects, for
example. The approach also holds promise for more complex problems 
 where small subsystems demanding fully wave-based 
analysis may coexist with larger structures necessitating ray-based 
approximations \cite{Maks11,Cha13}. In that context, the ability to perform fully wave-based
calculations within a formalism that is suitable also for large-scale ray
approximations will allow their efficient integration within hybrid methods.

Equations (\ref{singeq}) and (\ref{regeq}) lie at the heart of
everything calculated in this paper. Their key advantage is in allowing
a very natural in-out decomposition to be made  while owing nothing to
semiclassical approximation. There remains  a
significant degree of freedom, however, in choosing the singular parts
of the Green operators used to construct the outgoing component. The
simplest ``primitive'' choice set out in Sec.~\ref{primsec} appears very natural from the standpoint
of ray-based calculations and gives the right leading-order behaviour
straightforwardly (see \ref{Appsc}). It is singular, however, around criticially
incident waves arriving tangentially at the boundary and fails to
define appropriately decaying outgoing components in evanescent regimes (see 
Fig.~\ref{psiprimfigs} and the surrounding discussion in
Sec.~\ref{evansec}). We have therefore suggested in Sec.~\ref{regsec}
a second approach appropriate to domains
with analytic boundaries. Here, we have used asymptotic analysis of
the boundary integrals to motivate a decomposition based on integration
contours on the complexified boundary. This approach is at once able
to provide semiclassical approximation beyond all orders and amenable to exact 
analysis. It has been shown to work well in providing a solution of
model problems with circular geometry in Secs.~\ref{egsec} and \ref{sec:dielcav}.

Three outstanding problems remain which deserve further 
investigation. First, the applications so far carried out have been 
to the special case of a circular domain for which the shift operator 
can be fully determined analytically. The regular decomposition is
also applicable  to
more general (analytic) domains, but will require some numerical 
intervention in such cases to implement fully. Efficient means of
performing steps such as the operator inversions in (\ref{G1bis}) therefore
need attention, as do practical approaches to truncation of the complex
contours in the presence of chord-length singularities expected for
generic boundaries. Second, while we hope that equations
(\ref{singeq}) and (\ref{regeq}) will eventually provide a basis also
for the treatment of nonconvex domains, or boundaries with corners,
these will respectively require more careful analysis of the nontrivial
operator inversion on the right hand side of (\ref{genscat}) and the isolation
of the singular parts of boundary integrals at corners. 
Third, there remains some degree of ``ugliness" in the treatment of the transition 
from open to closed modes in the discussion provided for the regularised
decomposition. Although the regularised components do not diverge around
criticality as do their primitive counterparts, they do still 
undergo a discontinuous change. This is due to a coalescence and switching
of identities of incoming and outgoing components around critical 
reflection and may therefore be unavoidable, but one would ideally
manage the transition in a smoother way. 

\ack{SCC acknowledges support from EPSRC under grant number EP/F036574/1
and the hospitality of the MPIPKS, Dresden, where part of this work was completed 
while participating in the Advanced Study Group
\textit{Dynamical Tunneling}. HBH acknowledges support 
by the Libyan Ministry of Education. Further support by the EU (FP7 IAPP grant MIDEA)
is gratefully acknowledged. }

\appendix
\section{Green operator on a line}
\label{SingpartG}

The extraction of the singular part $\Ghat_0^\sing$ 
of the Green operator plays a central role in the
decomposition of the boundary functions on which we base
the transfer-operator representation of the boundary integral 
equations. In the case of the primitive decomposition described in
Sec.~\ref{primsec}, 
we have defined $\Ghat_0^\sing$ so that it corresponds to
replacing the boundary locally by a straight line. This allows
the simple, compact representation  given in (\ref{defGsing}), which 
is derived in this appendix.

Let $\Ghat_0^\sing$ be defined by (\ref{defGsingint}).
Before taking the limit $\eps\to 0$ in this definition,
the operator takes the form of a convolution of the test function
$\mu(s)$ with a function whose Fourier transform is
\begin{eqnarray}
\tilde{G}_0^\sing(p) 
&=&\frac{\rmi}{4}\int_{-\infty}^{\infty}
\rme^{-\rmi kps}H_0^{(1)}(k\sqrt{s^2+\eps^2})\rmd s\nonumber\\
&=& \label{Hankeltrans}
\cases{\frac{\rmi}{2k\sqrt{1-p^2}}\rme^{\rmi k\eps \sqrt{1-p^2}}&if $p^2<1$\\
\frac{1}{2k\sqrt{p^2-1}}\rme^{-k\eps \sqrt{p^2-1}}&if $p^2>1$,} 
\end{eqnarray}
where the latter form is tabulated in section 6.677 of \cite{Gradshteyn}.
Taking the limit $\eps\to 0$,
the action of $\Ghat_0^\sing$ on the Fourier transform
$\chi(p)$ of $\mu(p)$ is such that
\[
\Ghat_0^\sing \chi(p) = \chi(p)\times
\cases{\frac{\rmi}{2k\sqrt{1-p^2}}&if $p^2<1$\\
\frac{1}{2k\sqrt{p^2-1}}&if $p^2>1$.} 
\]
This justifies the formal definition (\ref{defGsing}) of the singular 
part of the Green function, along with the integral version given in 
(\ref{oppn}). The branches of the square root obtained for $p^2>1$ 
also explain the route of the contour in (\ref{oppn}) around the 
branch points $p^2=1$.

We now derive the equivalent expression (\ref{defG1sing})
for $\Ghat_1^\sing$, beginning with its formal definition
\[
\Ghat_1^\sing\psi(s) = \lim_{\eps\to 0}
\frac{\rmi k}{4}\int_{-\infty}^\infty 
\frac{\eps H_0'\left(k\sqrt{\vert s-s'\vert^2+\eps^2}\right)
}{\sqrt{(s-s')^2+\eps^2}}
\psi(s')\d s'
\]
analogous to (\ref{defGsingint}). Note that applying this operator
amounts taking a derivative with respect to $\eps$ in (\ref{defGsingint})
before taking the limit. Therefore, its action on the Fourier transform
$\varphi(p)$ of $\psi(s)$ is obtained by multiplying it by the derivative
\begin{eqnarray}
\tilde{G}_1^\sing(p) 
&=& 
\cases{-\frac{1}{2}\rme^{\rmi k\eps \sqrt{1-p^2}}&if $p^2<1$\\
-\frac{1}{2}\rme^{-k\eps \sqrt{p^2-1}}&if $p^2>1$,} 
\end{eqnarray}
of (\ref{Hankeltrans}) and then letting $\eps\to 0$, after which
\[
\Ghat_1^\sing\varphi(p) = -\frac{1}{2}\varphi(p),
\]
which demonstrates (\ref{defG1sing}).

\section{Semiclassical approximation of the primitive decomposition}
\label{Appsc}
In this appendix we describe more explicitly some of the semiclassical
results claimed in the main text for the primitive approximation.

\begin{figure}[!htp]
\centering
\includegraphics[scale=0.25]{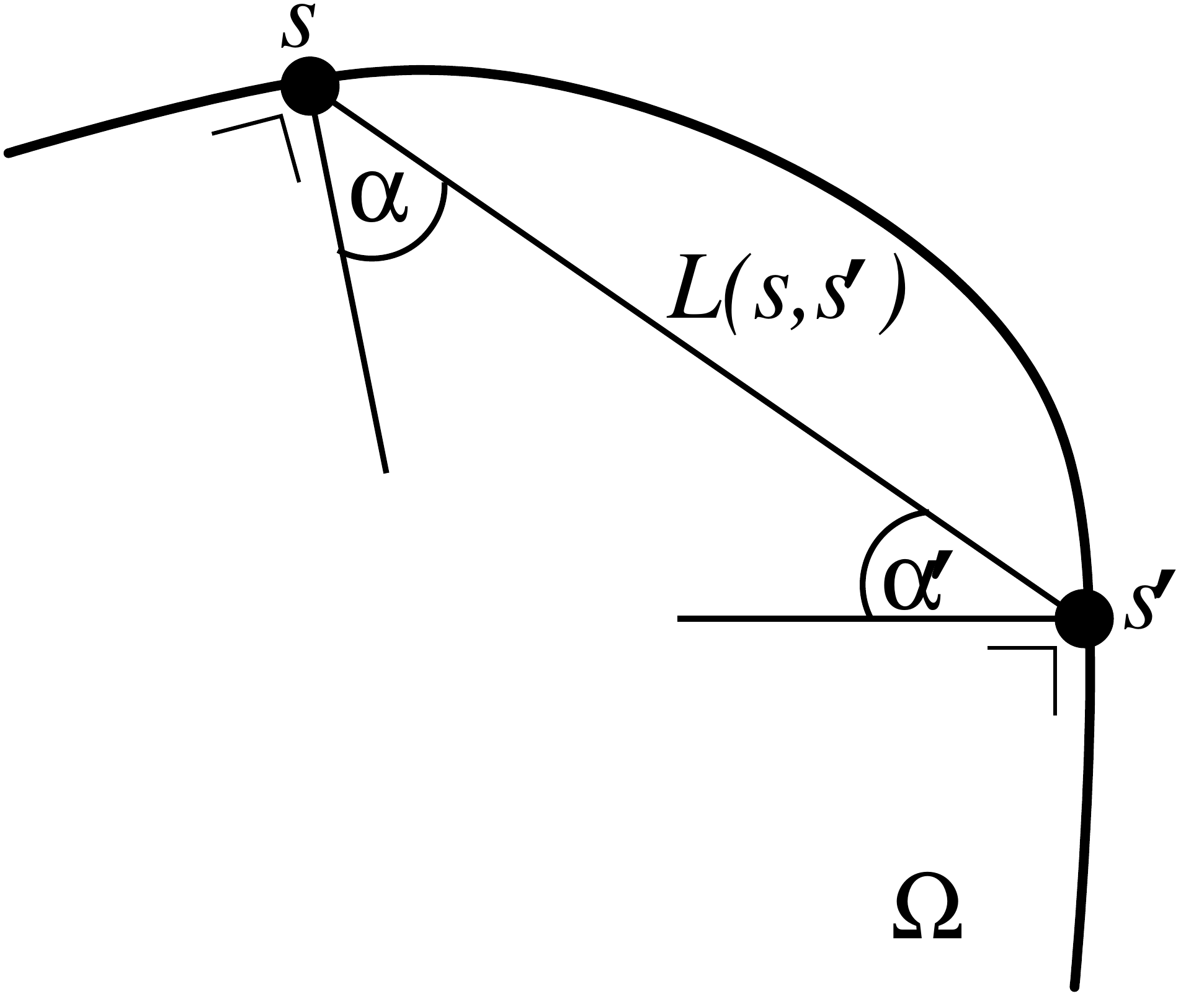}
\caption{The notation used for chord length and incidence angles 
is illustrated for the convex case.
}
\label{chord}
\end{figure}

To describe the primitive decomposition semiclassically, we begin with
the following rewriting of (\ref{defT0})
\[
\Shat_0 = -2\rmi k \Gh_0\pnhat-\Ih.
\]
We next note that applying the operator $\pnhat$ to a propagating 
boundary function of WKB form with local tangential momentum 
$p=\sin\alpha$ amounts at leading order to multiplication by 
$|\cos\alpha|$. Here $\alpha$ denotes the angle of incidence 
of the associated ray, defined with respect to a normal pointing \textit{into}
the domain $\Omega$ and we note that real chords on convex billiards
satisfy $\cos\alpha>0$. Away from the diagonal $s=s'$, where
the singular contribution from $\Ih$ can be neglected, the 
Green function can be approximated by
\[
G_0(s,s';k) \simeq 
\frac{\rmi}{2 k}\sqrt{\frac{k}{2\pi\rmi}
\frac{1}{\cos\alpha}
\dfdxdy{L}{s}{s'}\frac{1}{\cos\alpha'}}\;\e^{\rmi kL(s,s')}.
\]
For convex cavities, we find then that $\Shat_0$ can be represented at 
leading order by the kernel
\begin{equation}\label{defS0sc}
S_0(s,s')\simeq 
\sqrt{\frac{k}{2\pi\rmi}\frac{1}{\cos\alpha}\dfdxdy{L}{s}{s'}\cos\alpha'}
\;\e^{\rmi kL(s,s')} ,
\end{equation}
where $\alpha$ and $\alpha'$ are the chord's angles of incidence 
defined in Fig. \ref{chord}. One easily finds on using
$\dydxh{L}{n'}=\cos\alpha'$ that a corresponding 
calculation of the kernel $S_1(s,s')$ for $\Shat_1$ gives the same 
leading form.  This justifies (\ref{Ssequal}) for real rays in 
convex domains. 

The situation is different, however, for nonconvex domains or for
closed or evanescent modes represented by complex rays. Consider first the
case of real rays in nonconvex domains. In this case one
encounters ''ghost orbits'' which pass through the exterior of
$\Omega$ on the way from $s'$ to $s$. In particular, chords which
immediately leave $s'$ towards the exterior will be such
that $\dydxh{L}{n'}=\cos\alpha'<1$, in which case the kernels of
$\Shat_0$ and $\Shat_1$ have opposite sign at leading order:
\begin{equation}\label{cancelnot}
S_0(s,s')\approx -S_1(s,s').
\end{equation}
Note that ghost orbits can be shown to cancel systematically to all
orders in the full secular equation \cite{ghosts}, but their presence 
significantly complicates the present analysis by preventing a simple
evaluation of the operator inversion in (\ref{defT}). More generally
they prevent us from imposing (\ref{G1bis}) in a simple way. This
is not surprising as we expect the corresponding
\textit{external} problem, discussed at the end of Sec.~\ref{unishift},
to have nontrivial incoming components in nonconvex problems.

We state without proof that complex chords corresponding to evanescent
wave propagation beyond criticality also satisfy
(\ref{cancelnot}). This is reflected also in the vanishing of the
right hand side of (\ref{Sprim}) at leading order when $m^2>(kR)^2$.
The result is that $\psi_-=\Shat\psi_+$ is in general smaller than
$\psi_+$ in the closed region of momentum representation, although,
as evidenced by Fig.~\ref{psiprimfigs}, not exponentially so as the 
cancellation of the right hand side of (\ref{defT}) only occurs at
leading order and nontrivial higher-order contributions remain.

We remark finally that
a more explicit comparison with standard representations of 
the transfer operator can be made if the incoming and outgoing boundary 
waves are represented by $\Psi_\pm$ defined in (\ref{defPsiprim})
instead of $\psi_\pm$ and, replacing (\ref{defT0})  
and (\ref{defT1}), correspondingly. 
The transformed shift operators $\Th_0$ and $\Th_1$ are then defined by
\begin{equation}\label{defTt0}
\Gh_0 = \frac{\rmi}{2k}\frac{1}
{\sqrt{\pnhat}}\left(\Ihat + \Th_0\right)\frac{1}{\sqrt{\pnhat}}
\end{equation}
and
\begin{equation}\label{defTt1}
\Gh_1 = -\frac{1}{2}\frac{1}{\sqrt{\pnhat}}
\left(\Ihat + \Th_1\right)\sqrt{\pnhat}.
\end{equation}
The direct analogue of (\ref{defT}) now yields a shift operator 
$\Th=\pnhat^{1/2}\Shat\pnhat^{-1/2}$ such that $\Psi_-=\Th\Psi_+$, 
and which can be represented by a kernel whose leading semiclassical 
approximation takes the standard form
\[
\T(s,s')\simeq \T_0(s,s')\simeq \T_1(s,s')\simeq 
\sqrt{\frac{k}{2\pi\rmi}\dfdxdy{L}{s}{s'}}\;\e^{\rmi kL(s,s')} ,
\]
given in \cite{Bog}. 

\section{Alternative treatment of evanescent components}\label{Appalt}

The treatment of evanescent components within the regularised
decomposition has been described primarily in this paper while assuming 
that the endpoint contour component $\Gamma_0$ has been chosen to pass to
the right of the saddle in Fig.~\ref{contoursfig}(b). This has the
advantage of simplifying the presentation of formal results by
allowing, for example, condition (\ref{G1bis}) to hold at leading
order semiclassically over both the open and closed subspaces.

\begin{figure}[!htp]
\centering
\includegraphics[scale=0.19]{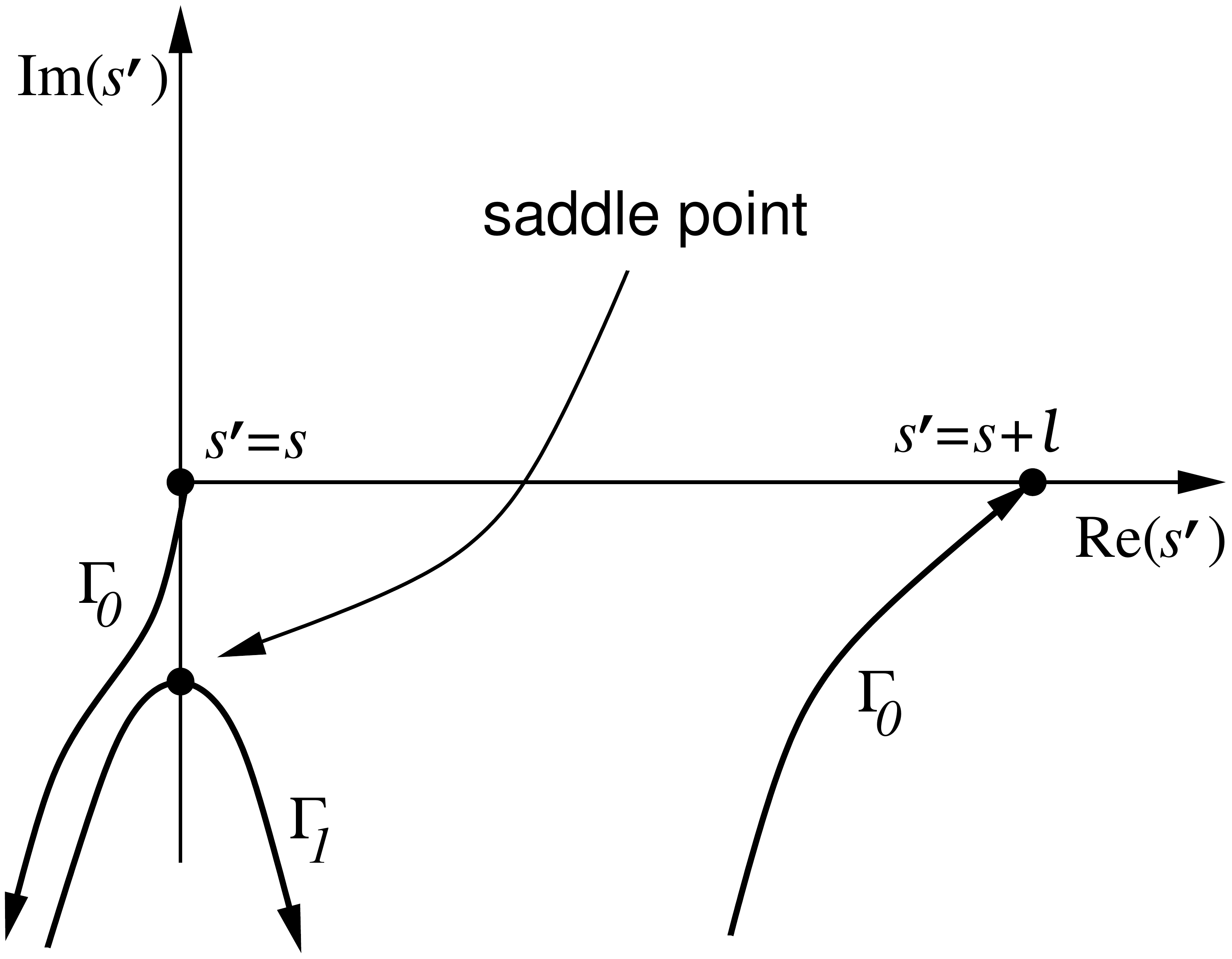}
\caption{The decomposition of contours illustrated here provides an
  alternative to the convention in Fig~\ref{contoursfig}(b).
}
\label{contoursbisfig}
\end{figure}

This is not the only possible choice of $\Gamma_0$, however. As
described in Sec.~\ref{regsec}, the saddle-point contribution to (\ref{singcont})
undergoes a Stokes transition around real values of $p^2>1$ so that it is a priori equally valid
for $\Gamma_0$ to pass to the other side of the saddle, as shown
schematically in Fig~\ref{contoursbisfig}. Choosing this alternative
route provides us with different forms for 
$\Gh_{0,1}^\sing$ and
$\Gh_{0,1}^\reg$ on the closed subspace, which we describe in this section for the case of 
the circle: note that, because $\Gamma_1$ is now nonempty, the
corresponding components of $\Gh_{0,1}^\reg$ are nonzero.
 We emphasise that, throughout this appendix, the
restrictions of $\Gh_{0,1}^\sing$ and
$\Gh_{0,1}^\reg$ to the open subspace are unaffected by this choice.

We find for this alternative contour decomposition that
\[
\left(\Gh_0^\sing\right)_{mm} = \frac{\rmi}{2k}\left(\rmi\pi
  zY_m(z)J_m(z)-\pi zJ_m(z)^2\right)
\]  
and 
\[
\left(\Gh_0^\reg\right)_{mm} = \frac{\rmi}{2k}\left(2\pi zJ_m(z)^2\right)
\] 
in the case of the circle, while
\[
\left(\Gh_1^\sing\right)_{mm} = -\frac{1}{2}\left(\pi z
  Y_m(z)J_m(z)+\pi\rmi zJ_m'(z)J_m(z)\right)
\] 
and 
\[
\left(\Gh_1^\reg\right)_{mm} = -\frac{1}{2}\left(-2\pi\rmi zJ_m'(z)J_m(z)\right)
\]
(all for $|m|>z$). This means in particular that condition
(\ref{G1bis}), does not automatically hold in this convention, since
\[
(\Shat_0)_{mm}\equiv
\left(\Gh_0^\reg \left(\Gh_0^\sing\right)^{-1}\right)_{mm} =
\frac{2J_m(z)}{\rmi Y_m(z)-J_m(z)},
\] 
while
\[
(\Shat_1)_{mm}\equiv
\left(\Gh_1^\reg \left(\Gh_1^\sing\right)^{-1}\right)_{mm} =
\frac{2J_m'(z)}{\rmi Y_m'(z)-J_m'(z)}
\]
(again, all for $|m|>z$). However, we do still find that considerable
simplification occurs in the shift operator itself, because the closed
components now satisfy
\begin{equation}\label{moremagic}
\left(\Gh_1^\reg-\Gh_0^\reg\left(\Gh_0^\sing\right)^{-1}\Gh_1^\sing\right)_{mm}
=\left(\Gh_0^\reg\left(\Gh_0^\sing\right)^{-1}\right)_{mm}
\end{equation}
(this identity is not immediately obvious but follows from
manipulation of the forms given above for 
$\Gh_{0,1}^\sing$ and
$\Gh_{0,1}^\reg$ and use of the Wronskian identities for Bessel functions).
In particular, we find from (\ref{genscat}) that the shift operator of the
interior problem still vanishes identically for closed modes $|m|>z$:
\[
\Shat^\inner = 0.
\]
The solution of the interior problem is then not qualitatively changed
by adopting this alternative convention and we simply find that, beyond
leading order, the construction of the shift operator is complicated
somewhat in the closed subspace.

On the other hand, the shift operator for the exterior problem
\textit{is} changed qualitatively. According to (\ref{genscatext}), 
the closed components of this operator then satisfy
\[
(\Shat^\ext)_{mm}=(\Shat_0)_{mm}= \frac{2J_m(z)}{\rmi Y_m(z)-J_m(z)},
\]
so that, although the external shift operator is exponentially
small ($J_m(z)\ll Y_m(x)$ in the evanescent regime), it does not vanish.
A convention in which the shift operator for the exterior
problem is exponentially small but nonvanishing may be an attractive
way of treating tunnelling effects directly. For convex domains,
waves leaving the boundary evanescently towards the exterior may
encounter, and be reflected at, caustics some distance from the 
boundary before returning to it. In the circular case such caustics 
are encountered at a radius $kr=|m|>kR$ where the outgoing Hankel
function undergoes a Stokes transition.
In the language of ray asymptotics,
although a real ray leaving the boundary never returns, a contributing complex
ray may (see \cite{usell} for an explicit discussion of such exterior
complex rays in the context of tunnelling emission from dielectric 
cavities).
In the convention used in the main text, such returning
evanescent waves are simply included in $\varphi_+$. The alternative
here allows such returning evanescent waves to contribute to a nonvanishing
component $\varphi_-$ and to be treated separately.

One final comment is that yet another convention is obtained by
defining $\Gh_{0,1}^\sing$ as
a weighted mean of the singular parts defined in this appendix and
those used in the main text. Although space does not allow us to
present details here, it is found that (\ref{moremagic}) still holds
in such conventions, as do the identities $(\Shat^\inner)_{mm}=0$ and 
$(\Shat^\ext)_{mm}=(\Shat_0)_{mm}$. Furthermore, a
symmetrically-weighted mean leads, for example, to operators $\Dhat_0$
and $\Dhat_1$ having purely imaginary components in the closed subspace.

\section{Mapping boundary solutions to the interior}\label{Appint}
In this appendix we summarise how the solution 
inside the domain can be determined in terms of the in-out boundary 
solutions. Let
\begin{eqnarray}
\label{defGG0}(\GGh_0\mu) (\x)&=&
\int_{\partial\Omega}G_0(\x,s';k)\mu(s')\d s'\\[3pt]
\label{defGG1}(\GGh_1\psi)(\x)&=&
\int_{\partial\Omega}\dydxv{G_0(\x,s';k)}{n'}\psi(s')\d s'
\end{eqnarray}
extend the Green operators defined by (\ref{defG0}) and
(\ref{defG1}) so that they map boundary functions to functions defined on the
interior $\Omega$. Then the interior solution $\psi(\x)$ can be
written formally
\[
\psi(\x) = \GGh_0\mu-\GGh_1\psi.
\]
Written in terms of the in-out components, this becomes (after some manipulation)
\[
\psi(\x) = 
\left(\rmi k\GGh_0\Dhat_0^{-1}(\Ihat+\rmi\Dhat_1)-\GGh_1\right)\psi_-
-\left(\rmi k\GGh_0\Dhat_0^{-1}(\Ihat-\rmi\Dhat_1)+\GGh_1\right)\psi_+.
\]
If the extended Green operators satisfy the following generalisation of
condition (\ref{G1bis})
\begin{equation}\label{G1bisbis}
\GGh_0\left(\Ghat_0^\sing\right)^{-1} = \GGh_1\left(\Ghat_1^\sing\right)^{-1},
\end{equation}
then the contribution from $\psi_-$ drops out of this equation after using the 
relations (\ref{D0}), (\ref{D1}) and the interior solution is given by the simplified 
expression
\[
\psi(\x) = -2\rmi k\GGh_0\Dhat_0^{-1}\psi_+.
\]
Condition (\ref{G1bisbis}) holds at leading order semiclassically for
all of the decompositions examined in this paper and can be shown to 
hold exactly in the case of the circle. How accurately it holds for
more general domains is a topic of investigation but it certainly
forms an appropriate starting point for semiclassical approximation.

\begin{figure}[!htp]
\centering
\includegraphics[scale=0.25]{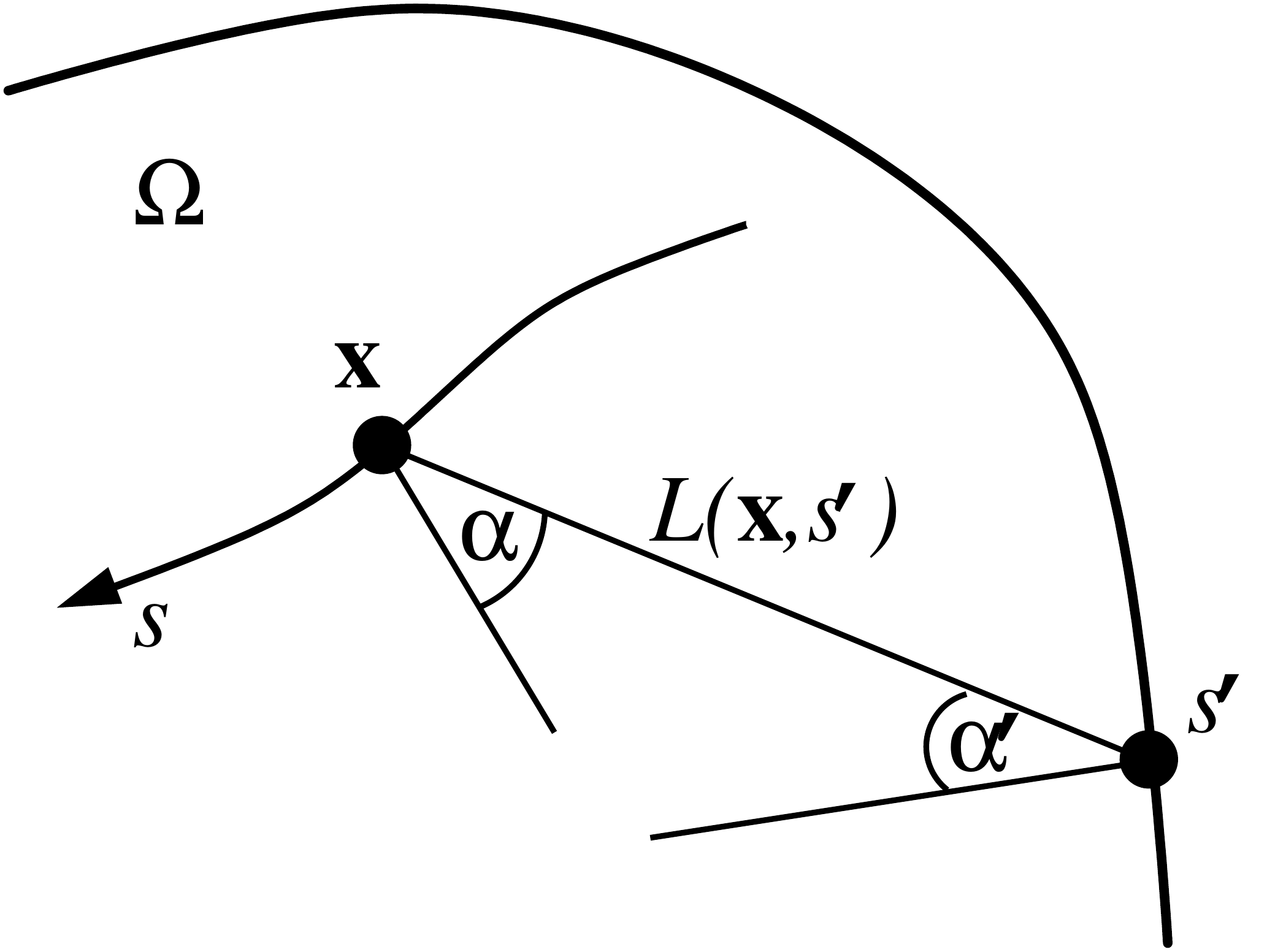}
\caption{The notation used in equation (\ref{defTT}) is illustrated.  
}
\label{interiorfig}
\end{figure}

Within semiclassical approximation one may therefore write
\[
\psi(\x)= \int_{\partial\Omega} \TT(\x,s')\psi_+(s')\d s',
\]
where
\[
\TT(\x,s') \simeq-2\rmi k G_0(\x,s';k)\cos\alpha'
\] 
and $\alpha'$ is the angle at which the line from $s'$ to $\x$
leaves the boundary. If we place $\x$ on a section parametrised by
arclength $s$ and such that the line from $s'$ arrives with incidence
angle $\alpha$, this can be approximated more explicitly by the form
\begin{equation}\label{defTT}
\TT(\x,s')\simeq 
\sqrt{\frac{k}{2\pi\rmi}\frac{1}{\cos\alpha}\dfdxdy{L}{s}{s'}\cos\alpha'}
\;\e^{\rmi kL(\x,s')} ,
\end{equation}
generalising (\ref{defS0sc}). We note finally that an analogous
boundary integral provides the solution $\varphi(\x)$ of the exterior 
problem in terms of the component $\varphi_+(s)$ leaving the boundary
towards infinity, except that the angles of incidence are defined
relative to the outward pointing normal in that case: in the
dielectric problem, $\alpha'$ is replaced by the angle of refraction 
$\beta'$.

\section*{References}
\bibliography{screferences}

\end{document}

%% file: macroz.tex
\def\braket#1#2{ \langle{#1}|{#2}\rangle }
\def\opket#1#2 {  {#1} |{#2}\rangle }

\def\d{{\rm d}}

\def\dfdxdy#1#2#3{  {   {\partial^2 #1} \over
            {\partial\mathstrut{#2}\partial\mathstrut{#3}} } }

\def\dydxh#1#2{ \partial{#1} / \partial {#2} }
\def\dydxv#1#2{ {{\partial #1} \over {\partial #2}} }

\def\det{{\rm det}\,}

\def\e{{\rm e}}

\def\eps{\varepsilon}

\def\ext{{\rm ext}}

\def\f0{\bar{f}}

\def\Fhat{\hat{F}}

\def\Gh{\hat{G}}
\def\g0{\bar{g}}

\def\h0{\bar{h}}
\def\H{{\cal H}}
\def\hh{\hat{h}}

\def\Ih{\hat{I}}

\def\inner{{\rm int}}

\def\K2{{\cal K}}
\def\ket#1{ |{#1}\rangle }

\def\max{{\rm max}}

\def\min{{\rm min}}

\def\phat{\hat{p}}

\def\prim{{\rm prim}}

\def\rmi{{\rm i}}
\def\rh{\hat{r}}

\def\T{{\cal T}}
\def\Th{{\hat{\cal T}}}

\def\Th{\hat{\T}}

\def\VV{{V}}

\def\V0{\bar{\VV}}
\def\WW{{W}}

\def\W0{\bar{\WW}}

\def\x{{\bf x}}

\def\spinor#1#2{\left( \begin{array}{c} {#1}\\[9pt]
			                {#2} \end{array} \right)}

\def\sbar{\big/}
\def\dbar{ {\,\sbar\!\sbar} }
\def\dbarr{ {\,\sbar} }

\def\opsket#1#2#3 {  {#1} \dbar_{#2}{#3}\rangle }

\def\opssket#1#2#3 {  {#1} \dbarr_{#2}{#3}\rangle }